\documentclass[journal]{IEEEtran}
\usepackage[cmex10]{amsmath}
\usepackage{algorithmic}
\usepackage{amssymb}
\usepackage{bm}
\usepackage{stmaryrd}  
\usepackage{amsfonts}
\usepackage{epsfig}

\usepackage{latexsym}
\usepackage{psfrag}
\usepackage{bm}
\usepackage{xspace}
\usepackage{graphicx}

\usepackage{watermark}

\usepackage{amsthm} 
\usepackage{wrapfig}
\usepackage{latexsym}
\usepackage{multicol}
\usepackage{float}
\usepackage{cite}
\usepackage{subfigure}
\usepackage{movie15,hyperref}

\newcommand{\ignore}[1]{}

\newcommand{\supp}{\operatorname{supp}}

\newcommand{\matr}[1]{\mathbf{#1}}
\newcommand{\vect}[1]{\mathbf{#1}}
\newcommand{\code}[1]{\mathcal{#1}}
\newcommand{\cC}{\mathcal{C}}
\newcommand{\set}[1]{\mathcal{#1}}

\newcommand{\Q}{\mathbb{Q}}

\newcommand{\tr}{\mathsf{T}}

\newcommand{\codeCQC}[1]{\code{C}_{\mathrm{QC}}^{(r)}}

\newcommand{\defeq}{\triangleq}

\newcommand{\vomega}{\boldsymbol{\omega}}

\newcommand{\onenorm}[1]{\lVert #1 \rVert_{\ell^1}}
\newcommand{\twonorm}[1]{\lVert #1 \rVert_{\ell^2}}

\newcommand{\Z}{\mathbb{Z}}

\renewcommand{\leq}{\leqslant}
\renewcommand{\geq}{\geqslant}

\newcommand{\setS}{\set{S}}

\newtheorem{lemma}{Lemma}
\newtheorem{theorem}[lemma]{Theorem}

\newtheorem{corollary}[lemma]{Corollary}
\newtheorem{conjecture}[lemma]{Conjecture}

\theoremstyle{plain}

\newtheorem{PreDefinition}[lemma]{{\textbf{Definition}}}
  \newenvironment{definition}%
    {\begin{PreDefinition}}{\hfill$\square$\end{PreDefinition}}

\theoremstyle{plain}

\newtheorem{PreRemark}[lemma]{{\textbf{Remark}}}
  \newenvironment{remark}%
    {\begin{PreRemark}\upshape}{\hfill$\square$\end{PreRemark}}

\newtheorem{PreExample}[lemma]{{\textbf{Example}}}
  \newenvironment{example}%
    {\begin{PreExample}\upshape}{\hfill$\square$\end{PreExample}}

\newcommand{\fch}[2]{\set{#1}(\matr{#2})}

\newcommand{\wps}{w_{\mathrm{p}}}
\newcommand{\wpsAWGNC}{w_{\mathrm{p}}^{\mathrm{AWGNC}}}

\newcommand{\wpsmin}{w_{\mathrm{p}}^{\mathrm{min}}}

\newcommand{\wpsminh}[1]{\wpsmin(\matr{#1})}

\newcommand{\wH}{w_{\mathrm{H}}}

\newcommand{\zero}{{\mathbf 0}} 
\newcommand{\Ftwo}{{{\mathbb F}}_{\!2}}

\newcommand{\perm}{\operatorname{perm}}

\newcommand{\permZ}{{\perm}_{\Z}}

 \newcommand{\Real}{\mathbb{R}}

\newcounter{mytempeqcounter}

\newcommand{\bigformulatop}[2]{%
  \begin{figure*}[!t]
    \normalsize
    \setcounter{mytempeqcounter}{\value{equation}}
    \setcounter{equation}{#1}
    #2

    \setcounter{equation}{\value{mytempeqcounter}}
    \hrulefill
    \vspace*{4pt}
  \end{figure*}
}

\title{Pseudocodewords from Bethe Permanents}
\author{Roxana Smarandache,~\IEEEmembership{Member,~IEEE}
  \thanks{Supported by NSF Grant   TF-0830608, CIF-1252788.}
  \thanks{R.~Smarandache is with the Departments of Mathematics and Electrical Engineering at the University of Notre Dame, Notre Dame, IN 46556, USA  (e-mail:
    rsmarand@nd.edu).}
}

\begin{document}
\maketitle

\begin{abstract}
 It was recently conjectured that  a vector with components equal to the Bethe permanent of certain submatrices of a parity-check matrix  is a pseudocodeword. In this paper we prove a stronger version of this conjecture for some important cases and investigate the families of  pseudocodewords obtained by using the Bethe  permanent. We also highlight some interesting properties of the permanent of block matrices and their effects on pseudocodewords.  
  \end{abstract}

\IEEEpeerreviewmaketitle

\section{Introduction} 
\IEEEPARstart{I}n~\cite{MacKay:Davey:01:1}, a  simple
technique is presented for upper bounding the  minimum Hamming distance of a binary
linear  code  that  is  described  by  an $m\times n$  parity-check  matrix $\matr{H}$. This is done 
based on  explicitly constructing codewords with components equal to $\Ftwo$-determinants 
of some $m\times m$  submatrices  of $\matr{H}$. Subsequently, this technique was extended and refined in~\cite{Smarandache:Vontobel:04:1, Smarandache:Vontobel:06:2:subm, Smarandache:Vontobel:09, 2012arXiv1201.2386B, 2011arXiv1111.0711W,  2012arXiv1210.3906P, Park:11} in the case of
quasi-cyclic binary linear codes. By computing those determinant components over the ring of integers $\Z$ 
instead of over the binary field $\Ftwo$ (and taking their absolute value) it was shown that the resulting integer vectors are pseudocodewords, called {\em absdet-pseudocodewords},  i.e.,  vectors that lie in the fundamental cone of the parity-check matrix of the code \cite{Feldman:03:1, Feldman:Wainwright:Karger:05:1}. In addition,  in \cite{Smarandache:Vontobel:09}, a closely related class of pseudocodewords called {\em perm-pseudocodewords} was defined, obtained by taking the vector components to be equal to the $\Z$-permanent of some $m\times m$ submatrices of $\matr{H}$ instead of the determinant. Pseudocodewords are  significant objects in iterative decoding and linear programming decoding. They  were first  introduced in~\cite{Wiberg:96} as culprits that prevent the iterative decoding algorithm to converge to a codeword  and further developed and studied  extensively (see  \url{www.pseudocodewords.info/} for a complete  list of papers on this and other related topics).   Although the above mentioned papers \cite{MacKay:Davey:01:1, Smarandache:Vontobel:04:1, Smarandache:Vontobel:06:2:subm, Smarandache:Vontobel:09, 2012arXiv1201.2386B, 2011arXiv1111.0711W,  2012arXiv1210.3906P, Park:11} give only upper bounds on the minimum distance,  they also contain useful guidelines for selecting suitable protographs with optimal minimum distance among all protographs.   Similarly, although the class of absdet- and perm-pseudocodewords 
provides upper bounds on the  minimum pseudo-weight, they represent good candidates for guiding the design of
low-density parity-check matrices. 

Related to the construction of  perm-pseudocodewords, Vontobel introduced in \cite[Sec.~IX]{Vontobel:11:3:subm} a similar vector but having components equal to the Bethe permanent of some $m\times m$ submatrices of a matrix $\matr{H}$ instead of the regular permanent, and conjectured that this vector is a pseudocodeword.
The term {\em Bethe permanent} was first used  by Vontobel in \cite{Vontobel:11:3:subm}, but the concept was introduced earlier in \cite{2009arXiv0908.1769H,2011arXiv1108.0065C, 2008arXiv0806.1199C},   to denote the approximation of a permanent of a non-negative matrix, i.e., of a matrix containing only non-negative real entries, by solving a certain Bethe free energy minimization problem.  In his paper \cite{Vontobel:11:3:subm}, Vontobel provided some reasons why the approximation works well, by showing that the Bethe free energy is a convex function and that the sum-product algorithm finds its minimum efficiently. 
Therefore, the Bethe permanent can be computed efficiently (i.e., in polynomial time) and so can be the Bethe perm-pseudocodeword based on  a set $\setS$  of some given column selection of the parity-check matrix. This is not the case for the perm-pseudocodeword. Therefore, the set of Bethe perm-pseudocodewords, together with that of absdet-pseudocodewords, also efficiently computed due to the polynomial-time computation of the determinant, constitute useful objects in determining  upper  bounds on the minimum pseudo-weight and  guiding the design of low-density parity-check matrices.   
In this paper we give four equivalent statements of the conjecture, prove a stronger version of the above mentioned conjecture in some cases  and discuss certain properties of Bethe permanents and their relationships to pseudocodewords.

The literature on permanents and adjacent areas of counting perfect matchings, counting 0-1 matrices with specified row and column sums, etc., is vast. Most relevant to this paper are the paper \cite{Ryser:63}  in which one of the most efficient algorithms for computing the permanent is given that requires $\Theta(n\cdot 2^n)$ arithmetic operations (better than brute force but still exponential in the matrix size),\footnote{ The number of real additions and multiplications needed to compute the determinant is in $O(n^3)$.} the paper \cite{Valiant:79} that shows that, in terms of complexity classes, the computation of the permanent is in the complexity class \#P,\footnote{\#P is the set of the counting problems associated with the decision problems in the class NP; even the computation of the permanent of matrices that contain only zeros and ones is \#P-complete.} 
the very recent work \cite{2011arXiv1108.0065C} that studies the so-called fractional free energy functionals and resulting lower and upper bounds on the permanent of a non-negative matrix, the paper \cite{2011arXiv1106.2844G}  that  shows that  the permanent is lower bounded by the Bethe permanent $\perm_{\mathrm{B}}(\theta) \leq \perm(\theta)$  and  discusses conjectures on what
the constant $C$ is in the inequality $\perm(\theta) \leq C \cdot \perm_{\mathrm{B}} (\theta)$,\footnote{An alternative way of proving Vontobel's conjecture  would be to start with the inequalities involving the permanents and then to apply sharp upper and lower bounds on these
permanents based on the Bethe permanents, and show that the
inequalities still hold. However, so far the known Bethe permanent
based upper and lower bounds on the permanent are not sharp enough to allow this approach.}  
the paper \cite{C:Greenhill:2010}  on counting perfect matchings in random graph covers, and the paper \cite{Bayati:Nair:06}  on counting matchings in graphs with the help of the sum-product algorithm.

The remainder of the paper is structured as follows. In Sections \ref{motivation} and \ref{numerical},  we give two examples intended to be motivational. The examples offer simple illustrations of some of the objects we study in the paper and explain their importance. In Section~\ref{sec:introduction}, we list basic notations and definitions, provide the necessary background, formally define the class of  perm-pseudocodewords and Bethe permanent vectors, state the conjecture and give a few examples to better illustrate the new notions and the conjecture. The following Section \ref{sec:results} gives the main results of this paper. In Section \ref{equiv:conj}, we show how the conjecture can be simplified to include only matrices of a certain form for which only one inequality is needed and from this, how the conjecture is equivalent to a certain co(perm)factor expansion on a row of a square matrix.
 We discuss the rows of the parity-check matrix of degree 2 or lower  in Section \ref{sec:case:degree:2}. 
We prove  a stronger version of the conjecture for two special cases in Sections \ref{case:conj} and  \ref{case:conj2} and discuss the next case of interest in Section \ref{case:conj3}.  In Section  \ref{2-3-0-section} and   Section \ref{3-4-0-section} we compute  the Bethe permanent pseudocodewords exactly for $2\times 3$ and $3\times 4$ matrices and offer some remarks and examples.  We conclude the paper in Section~\ref{sec:conclusions} with a summary of the solved and unsolved cases of the conjecture and offer some suggestions for future directions in attempting the open problems. In the Appendix \ref{proof:lemma} we included a proof of an important  lemma used in Section \ref{case:conj}. We give  expressions in closed form for some  permanents that are used throughout the paper in Appendices \ref{exact:q2} and \ref{exact:t3}. In Appendix \ref{lemmas} we prove three lemmas used in Appendices \ref{exact:q2} and \ref{exact:t3}.

\subsection{Motivational Example} \label{motivation}
 In this and the next sections we offer two examples, one more general and one numerical, which provide some motivational ground for the remainder of this paper. 
\begin{example}\label{example-motivation1} In \cite{MacKay:Davey:01:1}, the authors point out a weakness of the LDPC codes described by parity-check matrices $\matr{H}=(\matr{H}_{ij}) \in \Ftwo^{mM \times nM}$ with all  sub-matrices $\matr{H}_{ij} \in \Ftwo^{M \times M}$ commuting permutation matrices, i.e., matrices satisfying $\matr{H}_{ij}\matr{H}_{lk}=\matr{H}_{lk}\matr{H}_{ij}$, for all $i,j,l,k$. The codes associated with these matrices have minimum distance upper bounded by $(m+1)!$. In the proof,  the authors construct a codeword with weight at most $(m+1)!$, in the following way. For a set $\beta$ of $m+1$ indices in $[n]$, a matrix $\vect{D}_i$ 
is defined as the `determinant' modulo 2 of the $m\times m$ submatrix of the matrix $\matr{H}=(\matr{H}_{ij})$ with block-column indices  $\beta\setminus i$. For example, for $m=3$, $n=4$, matrix $\matr{H}$ of the form
\begin{align*}
  \matr{H}
    &= \begin{bmatrix}
         \matr{H}_{11}   & \matr{H}_{12}  & \matr{H}_{13}& \matr{H}_{14} \\
         \matr{H}_{21}   & \matr{H}_{22}  &\matr{H}_{23}  & \matr{H}_{2 4} \\
         \matr{H}_{31} & \matr{H}_{32} & \matr{H}_{33}&  \matr{H}_{34}
       \end{bmatrix},
\end{align*}
and $\beta=\{1,2,3,4\}$, we define  $\vect{D}_4\defeq \matr{H}_{11} \matr{H}_{22} \matr{H}_{33}+ \matr{H}_{12} \matr{H}_{23} \matr{H}_{31}+
\matr{H}_{13} \matr{H}_{21} \matr{H}_{32}+ \matr{H}_{13} \matr{H}_{22} \matr{H}_{31}+ \matr{H}_{12} \matr{H}_{21} \matr{H}_{33} + \matr{H}_{11} \matr{H}_{23} \matr{H}_{32}.$ The expressions $\vect{D}_1, \vect{D}_2$ and $\vect{D}_3$ are defined similarly.  Although  $\vect{D}_i$ are computed as `determinants', they are in fact  matrices of size $M\times M$.  Then, the vector $\vect{v}\defeq \left(\vect{D}_1\vect{x}, \vect{D}_3\vect{x}, \vect{D}_3\vect{x}, \vect{D}_4\vect{x}\right)$ is a codeword for the code described by $\matr{H}$, for any binary vector $\vect{x}$ of length $4M$.  Taking $\vect{x}$ of weight 1, we obtain the known bound $(3+1)!=(m+1)!$ on the minimum distance of protograph based LDPC codes with entries commuting matrices \cite{MacKay:Davey:01:1}. 
This method can be extended to any $m\times n$ matrix by considering vectors based on sets $\beta\subset [n]$ of size $m+1$, with components equal to 0 outside $\beta$.   

Note that the determinant $\vect{D}_i$ defined above is in fact a permanent in the free commutative ring $\cal R$ on the set of generators $\matr{H}_{ij}$.  
As matrices, each $\vect{D}_i$ has a 
$\Z$-permanent; 
instead of taking $\vect{D}_i\vect{x}$ which is the $i$th component of  the above vector $\vect{v}$, we can consider the vector 
\begin{align*} \vect{w}&\defeq\left[\perm_{\Z}(\vect{D}_1), \perm_{\Z}(\vect{D}_2), \perm_{\Z}(\vect{D}_3), \perm_{\Z}(\vect{D}_4) \right],\end{align*} 
with the $i$th component 
$\vect{w}_i$ equal to the 
permanent 
of the matrix $\vect{D}_i$.
This vector is obtained by applying two operators, first the $\mathcal R$-permanent operator  $\perm_{\mathcal R}$ on the $m\times m$ submatrices $\matr{H}_i$ of $\matr{H}$ to obtain $$\vect{D}_i=\perm_{\mathcal R}(\matr{H}_i)$$ and then the 
$\Z$-permanent operator $\permZ$ 
on the resulting matrices $\vect{D}_i$ to obtain 
  $$\vect{w}_i=\permZ(\vect{D}_i)=\permZ(\perm_{\cal R}(\matr{H}_i)).$$
 
 In addition, another  vector can be similarly considered. Let 
\begin{align*}\vect{\tilde{w}}\defeq&\left[\perm_{\Z}(\vect{H}_1), \perm_{\Z}(\vect{H}_2), \perm_{\Z}(\vect{H}_3), \perm_{\Z}(\vect{H}_4) \right], \end{align*}
with the $i$th component 
$\vect{\tilde{w}}_i$ equal to the 
permanent 
of the matrix $\vect{H}_i$ obtained from $\matr{H}$ by deleting the $i$th block column (i.e.,  with columns indexed by $\beta\setminus i$).

We need to mention that in general  $\tilde{\vect{w}}\neq \vect{w}$. 
So we have two different vectors related to the block parity-check matrix $\matr{H}$. 
 They are of length 4, which is the number of variable nodes in the protograph associated with the $3\times 4$  all-one matrix. The matrix $\matr{H}$ is a cover of this protograph.\footnote{We abuse the language and call a cover graph  of a given graph to also be a cover graph of the parity-check matrix of that given graph.}   Are these vectors pseudocodewords for the protograph, is there any useful information to be obtained from them? 
For simplicity in the following considerations, we will assume that all matrices $\matr{H}_{ij} \in \Ftwo^{M \times M}$ are non-zero.  It turns out that the vectors are not always pseudocodewords, however, taking the $M$th root of each component, the result is more likely to be a pseudocodeword.  In addition, when taking the $M$th root of the average of the permanents over  all covers, then the result is for sure a pseudocodeword.  This represents the Bethe permanent vector of $\theta$ based on the set $\beta=\{1,2,3,4\}$, where 
 \begin{align*}
  \theta
    &= \begin{bmatrix}
         1&1&1&1 \\1&1&1&1 \\1&1&1&1 
       \end{bmatrix}
\end{align*}
is the protomatrix of the protograph of $\matr{H}$. 
  This fact might not seem practical in this form. However,  it shows a connection between pseudocodewords associated with a matrix $\theta$ and the parity-check matrices of the cover graphs of that matrix, thus providing us with some  theoretical information about the mathematics of pseudocodewords. 
The practical aspect of this result is apparent when applying  the above procedure of computing the Bethe permanent vector based on a set $\beta$ of size $3M+1$ directly to the $3M\times 4M$ matrix $\matr{H}$, i.e., when we consider $\matr{H}$ to be the protomatrix. Any  set $\beta$ of size $3M+1$ of  columns of $\matr{H}$ yields a Bethe permanent vector based on that  set. Taking the nonzero minimum of their pseudo-weights, we obtain an upper bound on the minimum pseudo-weight of the matrix $\matr{H}$.   The vectors obtained in this manner are related to the perm-pseudocodewords based on a set $\beta$  of size $3M+1$ formally defined in the next section, in that instead of taking the permanent operator to compute the components, we use the Bethe permanent. As we mention in the introduction, computing the Bethe permanent is polynomial in time, so these vectors could be more useful in computing upper bounds on the  minimum pseudo-weight than the perm-pseudocodewords. 

\end{example}
\subsection{Numerical Example}\label{numerical}
 
 \begin{example} \label{example-motivation}
Let $M=3$, $P^s$, $s = 0, 1, 2$, be the  $s$-times cyclically
  left-shifted identity  matrices of size $M \times M$. Let $\matr{H}$ be a parity-check matrix based on the protomatrix $\theta$, with 
 \begin{align*} 
  \theta&  = \begin{bmatrix}
       1&1&1&1\\0&1&1&1\\0&1&1&1
         \end{bmatrix},\ \ \ \matr{H}
    = \begin{bmatrix}
       I  & I & I & I \\
           0 & I & P & P^2 \\
         0   & I & P^2 & P
         \end{bmatrix}.
\end{align*}
The $\vect{D}_i$ matrices are\footnote{We used that, for $M=3$,  $ P^4=P$.}
 \begin{align*}
\vect{D}_1=&  P^2+2P^2+ 2P^4+P^4=3P+ 3P^2\\  
\vect{D}_2=& P^2+P^4=P+P^2,  \vect{D}_3=  P+P^2,  \vect{D}_4=  P+P^2.
\end{align*}
 
 The binary vector $$\vect{v}\!\defeq \!\left(\vect{D}_1\vect{x}, \vect{D}_2\vect{x}, \vect{D}_3\vect{x}, \vect{D}_4\vect{x}\right)\!=\! \left( 0,1,1,\ 0,1,1,\ 0,1,1,\ 0,1,1\right)$$ gives  a codeword of length $12$ in the code represented by $\matr{H}$,  
where $\vect{x}$ denotes as above  a vector of weight 1. ($\vect{D}_i\vect{x}$ is then a column of the matrix  $\vect{D}_i$). 

Note that, by projecting  this vector onto $\Real^4$, we obtain  the pseudocodeword $(2/3,2/3,2/3,2/3)$ equivalent to the unscaled 
pseudocodeword $(1,1,1,1)$ of pseudo-weight 4, for the  protomatrix $\theta$ of $\matr{H}$ given above.  

The integer vector $\vect{w}$ is computed as follows
 \begin{align*}\vect{w}\defeq&\left[\perm_{\Z}( 3P+ 3P^2), \perm_{\Z}(P+P^2), \right.\\&\left. \perm_{\Z}( P+P^2), \perm_{\Z}(P+P^2) \right], \end{align*}
 and gives the vector $(54,2,2,2)$. 
 This is not a pseudocodeword for  $\theta$, 
 but
 $\vect{w}^{1/M} \defeq(54^{1/M},  2^{1/M},2^{1/M},2^{1/M}) =(54^{1/3},1^{1/3},1^{1/3},1^{1/3})=\sqrt[3]{2}(3,1,1,1)$  is a pseudocodeword equivalent to the unscaled pseudocodeword $(3,1,1,1)$ which has AWGNC-pseudo-weight $3$. This is lower than the pseudo-weight of the projection of the vector $\vect{v}$ obtained by the standard  method of projecting codewords in a cover graph of a graph of interest onto pseudocodewords for this graph.  

The vector  $\vect{\tilde{w}} $ is computed as follows
 \begin{align*}\vect{\tilde{w}}=&\left[\perm_{\Z}\begin{bmatrix}
         \matr{I} & \matr{I} & \matr{I} \\ 
         \matr{I}& P & P^2 \\
         \matr{I} & P^2 & P
         \end{bmatrix}, \perm_{\Z}  \begin{bmatrix}
    P& P^2 \\
 P^2 & P
         \end{bmatrix}, \right. \\ & \left. \perm_{\Z}\begin{bmatrix}
    \matr{I} & P^2 \\
 \matr{I}&P
         \end{bmatrix},\perm_{\Z}\begin{bmatrix}
    \matr{I} & P \\
 \matr{I} & P^2
         \end{bmatrix} \right]\end{align*} which gives the vector $\vect{\tilde{w}}=(48,2,2,2)$. 

This vector is not a pseudocodeword,
 but $\vect{\tilde{w}}^{1/M} \defeq(48^{1/M},  2^{1/M},2^{1/M},2^{1/M}) =(48^{1/3},1^{1/3},1^{1/3},1^{1/3})=\sqrt[3]{2}(2\sqrt[3]{3},1,1,1)$  is a pseudocodeword equivalent to the unscaled pseudocodeword $(2\sqrt[3]{3},1,1,1)$ which has AWGNC-pseudo-weight $3.58$.

Is this always true?  The answer is no, there are cases for which this does not happen. However, we show  that this fact  is true if we take the $M$th root of the  average of all possible liftings of $\theta$ instead of the $M$th root of the permanent of one single lifting. 
 
   \end{example}

The above examples are very simple examples (since we obtain pseudocodewords of length 4)  to explain the idea of Bethe permanents. It is applied to the matrix $\theta$, i.e., we obtain pseudocodewords for the protograph associated to $\theta$. However,  this same method can be similarly used to obtain pseudocodewords for the matrix $\matr{H}$.  We will revisit this example after the following  introductory chapter.


\section{Definitions, Vontobel's Conjecture and Examples} \label{sec:introduction}

Let $\Z$, $\Real$, and $\Ftwo$ be the ring of integers, the field of real
numbers, and the finite field of size $2$, respectively. Rows and columns of matrices and entries of vectors will be indexed starting at $1$. 
If $\matr{H}$ is some matrix and if $\alpha=\{i_1,  \dots, i_r\}$ and $\beta=\{j_1, \ldots, j_s\}$ are subsets of the row
and column index sets, respectively, then $\matr{H}_{\alpha,\beta}$ is the
sub-matrix of $\matr{H}$ that contains only the rows of $\matr{H}$ whose index
appears in the set $\alpha$ and only the columns of $\matr{H}$ whose index
appears in the set $\beta$. If $\alpha$ is the set of all row indices of
$\matr{H}$, we will simply write $\matr{H}_{\beta}$ instead of
$\matr{H}_{\alpha,\beta}$. Moreover, for any set of indices $\gamma$,  we will use the short-hand $\gamma
\setminus i$ for $\gamma\setminus \{ i \}$. 
We will also use the common notation $h_{ij}$  to denote the $(i,j)$th entry of a matrix $\matr{H}$ when there is no ambiguity in the indices
and $h_{i,j}$ when one of the two indices is not a simple digit, e.g., $h_{i,m-1}$.   We will use $P_{ij}$ for matrix block entries of a matrix, 
and $P_{ij,\beta}$ for a submatrix of $P_{ij}$ with column indexed by $\beta$, when there is no ambiguity, and $P_{i,j, \beta}$  when one of the two indices $i,j$ is not a simple digit, e.g., $P_{i,m-1,\beta}.$
 For an integer $M$, we will use the common notation $[M]\defeq \{ 1,
  \ldots, M\}$. For a set $\alpha$, $|\alpha|$ will denote the cardinality of $\alpha$ (the number of elements in the set $\alpha$). The set of all $M\times M$ permutation matrices  will be denoted by ${\mathcal P}_M$. As usual, the set of all permutations on the set $[m]$ is denoted by ${\mathcal S}_m$. 

  \begin{definition}

  Let $\theta = (\theta_{ij})$ be an $m \times m$-matrix over some commutative ring.
  Its determinant and permanent, respectively,  are defined to be \vspace{-1mm}
  \begin{align*}
    \det(\theta)
      &\defeq \sum_{\sigma \in {\mathcal S}_m}
           \operatorname{sgn}(\sigma)
           \prod_{i\in [m]}
           \theta_{i\sigma(i)} \; ,\:\\
    \perm(\theta)
      &\defeq \sum_{\sigma \in {\mathcal S}_m}
           \prod_{i\in [m]}
           \theta_{i\sigma(i)} \; , 
  \end{align*}
where $\operatorname{sgn}(\sigma)$ is the signature operator.  \end{definition}
In this paper, we consider only permanents over the integers.
\begin{definition}\label{def:pseudocodewords}
Let $\matr{H} = (h_{ij})$ be an $m\times n$ parity-check matrix of some binary
linear code. The fundamental cone $\fch{K}{H}$ of $\matr{H}$ is the set of all vectors
  $\vomega=(\omega_i) \in \Real^n$ that satisfy 
  \begin{alignat}{2}
    \omega_j
      &\geq 0 
      \ 
      &&\text{ for all $j \in [n]$} \; , 
         \label{eq:fund:cone:def:1} 
          \\
    \omega_j
      &\leq
          \sum_{j' \in \supp(\vect{R}_i) \setminus j} \!\!
            \omega_{j'}
      \ 
      &&\text{for all $i \in [m]$ \ and 
                $j \in \supp(\vect{R}_i)$} \; ,
          \label{eq:fund:cone:def:2}
  \end{alignat}
where $\vect{R}_i$ is the $i$th row vector of $\matr{H}$ and $\supp(\vect{R}_i)$ is its support (the positions 
where the vector is non-zero). 
  A vector $\vomega \in \fch{K}{H}$ is called a {\em pseudocodeword}~\cite{Vontobel:Koetter:05:1:subm}. Two pseudocodewords $\vomega, \vomega' \in
  \fch{K}{H}$ are said to be in the same equivalence class if there exists an $\alpha >
  0$ such that $\vomega = \alpha \cdot \vomega'$. In this case, we write $\vomega \propto \vomega'$. \end{definition}
There are two ways of characterizing pseudocodewords in a code $\cC$: as codewords in codes associated to covers of the Tanner graph
corresponding to $\cC$, or using the computationally easier {\em linear programming  (LP) approach}, which connects the presence of pseudocodewords
in message passing iterative decoding and LP decoding~\cite{Koetter:Vontobel:03:1, Vontobel:Koetter:05:1:subm}. Here we take  the latter approach.

In the same way as the minimum Hamming weight  is a very important
characteristic for a binary linear code under maximum-likelihood (ML) decoding, the minimum pseudo-weight
$\wpsminh{H}
    \defeq
       \min_{\vomega \in \set{P}(\matr{H}) \atop \vomega \neq \vect{0} }
         \wps(\vomega)$ is a very important characteristic for a binary linear code under message passing iterative 
decoding, especially for large signal-to-noise ratios (SNRs).
Pseudo-weights are
  channel-dependent; in this paper we assume  the binary-input AWGNC pseudo-weight of a pseudocodeword $\vomega \neq \vect{0}$
   defined as  ~\cite{ Wiberg:96,
    Forney:Koetter:Kschischang:Reznik:01:1, Koetter:Vontobel:03:1,
    Vontobel:Koetter:05:1:subm}
\begin{align*}
  \wpsAWGNC(\vomega)
    &\defeq \frac{\onenorm{\vomega}^2}
                 {\twonorm{\vomega}^2},
\end{align*}
where $\onenorm{\,\cdot\,}$ and $\twonorm{\,\cdot\,}$ are, respectively, the
$1$-norm and $2$-norm. Note that if $\vomega$ is a vector containing only
zeros and ones, then $\wpsAWGNC(\vomega) = \wH(\vomega)$.

  \begin{definition}
  \label{def:perm:vector:1} 
 Let $\code{C}$ be a  binary linear code described by a
  parity-check matrix $\matr{H}\in \Ftwo^{m \times n}$, $m < n$. For a
  size-$(m{+}1)$ subset $\beta$ of $[n]$ we define the perm-vector
  based on $\beta$ to be the vector $\vomega \in \Z^n$ with components\vspace{-2mm}
  \begin{align*}
    \omega_i
      &\defeq
         \begin{cases} 
           \perm\big( \matr{H}_{\beta \setminus i} \big) 
             & \text{if $i \in \beta$} \\
           0                                     
             & \text{otherwise} \; .
         \end{cases} \\[-1.2cm]
           \end{align*}
\end{definition}  
The permanent operator is taken over $\Z$. In \cite{Smarandache:Vontobel:09} it was shown that these vectors are in fact pseudocodewords. We state this here for easy reference  together with its proof. 
\begin{theorem} (from \cite{Smarandache:Vontobel:09}) \label{theorem:perm:vector:1}
 Let $\code{C}$ be a binary linear code described by the parity-check matrix
  $\matr{H}\in \Ftwo^{m \times n}$, $m < n$, and let $\beta$ be a
  size-$(m{+}1)$ subset of $[n]$. The perm-vector $\vomega$ based
  on $\beta$ is a pseudocodeword of $\matr{H}$.\end{theorem}
\begin{IEEEproof}
  We need to
  verify~\eqref{eq:fund:cone:def:1} and~\eqref{eq:fund:cone:def:2}. From
  Definition~\ref{def:perm:vector:1} it is clear that $\vomega$
  satisfies~\eqref{eq:fund:cone:def:1}. To show 
  that $\vomega$ satisfies~\eqref{eq:fund:cone:def:2}, we fix an $i \in
  [m]$ and a $j \in  \supp(\vect{R}_i)$. If $j \notin \beta$
  then $\omega_j = 0$ and~\eqref{eq:fund:cone:def:2} is clearly satisfied.
  If $j \in \beta$, then
  \begin{align*}
    \sum_{j' \in   \supp(\vect{R}_i)\setminus j}
     \!\!\! \!\!\!\!\omega_{j'}
      = &\!\!\sum_{j' \in  \supp(\vect{R}_i)\setminus j}
           h_{ij'}
           \omega_{j'} \\
      =&\!\! \sum_{j' \in \beta \setminus j}
           h_{ij'}
           \cdot
           \perm\big( \matr{H}_{\beta\setminus j'} \big)
         +\!\!\!\!
         \sum_{j' \in ([n] \setminus \beta) \setminus j}\!\!\!
           h_{ij'}
           \cdot
           0 \\
      =&\!\! \sum_{j' \in \beta \setminus j}
           h_{ij'} \!\!\!\!
           \sum_{j'' \in \beta \setminus j'}
             h_{ij''}
             \perm
               \big(
                 \matr{H}_{[m] \setminus i, \beta \setminus \{ j', j'' \}}
               \big) \\
   \overset{(*)}{\geq}& \!\!
         \sum_{j' \in \beta \setminus j}
           h_{ij'}
           h_{ij}
             \perm
               \big(
                 \matr{H}_{[m] \setminus i, \beta\setminus \{ j', j \}}
               \big)\\ 
                  =& \ h_{ij}
         \sum_{j' \in \beta \setminus j}
           h_{ij'}
           \perm
             \big(
               \matr{H}_{[m] \setminus i, \beta \setminus \{ j', j \}}
             \big) \\
      =&\  h_{ij}
         \perm
           \big(
             \matr{H}_{\beta \setminus j }
           \big)
       = h_{ij} \omega_j
       \overset{(**)}{=}
         \omega_j \; ,
  \end{align*}
  where at step $(*)$ we kept only the terms for which $j'' = j$, and where
  step $(**)$ follows from $j \in \supp(\vect{R}_i)$. Because $i \in
[m]$ and $j \in\supp(\vect{R}_i) $ were taken arbitrarily, it follows that $\vomega$
 satisfies~\eqref{eq:fund:cone:def:2}.
\end{IEEEproof}

\begin{example}\label{example:1}
  Consider the $[4,2,2]$ binary linear code $\code{C}$ based on the
  parity-check matrix $\matr{H} \defeq \begin{bmatrix} 1 & 1 & 1 & 0 \\
    0 & 1 & 1 & 1 \end{bmatrix}$, where $n = 4$ and $m = 2$. The following list contains the
  perm-vectors based on all possible subsets $\beta\subset [4]$  of size $m{+}1 = 3$: $(2, 1, 1, 0)$, $(1, 1, 0, 1)$, $(1, 0, 1, 1)$, $(0,
  1, 1, 2)$. It can be easily checked that these satisfy the inequalities of the fundamental cone above, as the theorem predicts. 
  They give an upper bound on the minimum pseudo-weight of $8/3$. 
\end{example}

The following combinatorial description of the Bethe permanent can be found in \cite{Vontobel:11:3:subm}. We use it here as a definition. 
\begin{definition} Let $\theta$ be a non-negative (with non-negative real entries) $m \times m$ matrix and  $M$ be a  positive integer.  
Let $\Psi_{m,n, M}$ be the set $$\Psi_{m,n,M}\defeq\mathcal{P}_M^{m\times n}=\{\matr{P} = (P_{ij})_{\tiny \begin{matrix}i\in[m]\\ j\in [n]\end{matrix}} ~|~P_{ij} \in \mathcal{P}_M\}.$$ If $m=n$, we will use  $\Psi_{m,M}\defeq \Psi_{m,n,M}.$

For a matrix $\matr{P}\in  \Psi_{m,M}$, the $\matr{P}$-lifting of $\theta$  is defined as the $mM\times mM $ matrix
$$\theta ^{\uparrow\matr{P}}\defeq \left(\begin{matrix} \theta_{11}P_{ 11}&\dots & \theta_{1m}P_{1m} \\ \vdots &&\vdots\\ \theta_{m1}P_{m1}&\ldots &\theta_{mm}P_{mm}\end{matrix}\right),$$
and the degree-$M$ Bethe permanent of $\theta$ is defined as 
$$ \perm_{\mathrm{B},M}(\theta)\defeq  \sqrt[M] {\big<\perm(\theta^{\uparrow\matr{P}})\big>_{\matr{P}\in \Psi_{m,M}}},$$
where the angular brackets represent the arithmetic average of $\perm(\theta^{\uparrow\matr{P}})$ over all $\matr{P} \in \Psi_{m,M}$. 

Then,  the Bethe permanent of $\theta$ is  defined as 
\begin{align*}\perm_{\mathrm{B}}(\theta)\defeq \limsup_{M\rightarrow \infty} \perm_{\mathrm{B},M}(\theta). \\[-1.3cm]\end{align*}
\end{definition}

\begin{remark}  Note that a $\matr{P}$-lifting of a matrix $\theta$ corresponds to an $M$-graph cover of the protograph (base graph) described by $\theta$.  Therefore we can consider  $\theta ^{\uparrow\matr{P}}$ to represent a protograph-based LDPC
code and $\theta$ to be its protomatrix (also called its base matrix or its mother matrix) \cite{thor03}.  \end{remark} 

\begin{definition}
  \label{def:perm:bethe:vector:1} 
 Let $\code{C}$ be a  binary linear code described by a
  parity-check matrix $\matr{H}\in \Ftwo^{m \times n}$, $m < n$. For a
  size-$(m{+}1)$ subset $\beta$ of $[n]$ we define the Bethe permanent vector
  based on $\beta$ to be the vector $\vomega_{\mathrm{B}} \in \Real^n$ with components
  \begin{align*}
    \omega_{\mathrm{B},i}
      &\defeq
         \begin{cases} 
           \perm_{\mathrm{B}}\big( \matr{H}_{\beta \setminus i} \big) 
             & \text{if $i \in \beta$} \\
           0                                     
             & \text{otherwise}\; .
         \end{cases} 
  \end{align*}

Similarly,  we define degree-$M$ Bethe  permanent vector
  based on $\beta$ to be the vector $\vomega_{\mathrm{B},M} \in \Real^n$ with components
  \begin{align*}
    \omega_{\mathrm{B},M,i}
      &\defeq
         \begin{cases} 
           \perm_{\mathrm{B},M}\big( \matr{H}_{\beta \setminus i} \big) 
             & \text{if $i \in \beta$} \\
           0                                     
             & \text{otherwise}\; .
         \end{cases} \\[-1.2cm]
  \end{align*}
\end{definition}

The following conjecture is stated in~\cite{Vontobel:11:3:subm}.
\begin{conjecture} [\!\cite{Vontobel:11:3:subm}] Let $\code{C}$ be a binary linear code described by
an $m\times n$  binary parity-check matrix $\matr{H}$, with $m<n$,  and let $\beta$ be a
size-$(m+1)$ subset of $[n]$. Then the Bethe permanent vector $\vomega_{\mathrm{B}}$  based on $\beta$  is a pseudocodeword of $\matr{H}$, i.e.,
$\vomega_{\mathrm{B}}\in \fch{K}{H}$.
\end{conjecture}

\begin{example}\label{example:11}
  Consider the $[4,2,2]$ binary linear code $\code{C}$ based on the
  parity-check matrix $\matr{H} \defeq \begin{bmatrix} 1 & 1 & 1 & 0 \\
    0 & 1 & 1 & 1 \end{bmatrix}$, where $n = 4$ and $m = 2$. The following list contains the degree-$M$ Bethe  permanent vectors based on all possible subsets $\beta\subset [4]$  of size $m{+}1 = 3$:  $((M+1)^{1/M}, 1, 1, 0))$, $(1, 1, 0, 1)$, $(1, 0, 1, 1)$, $(0, 1, 1, (M+1)^{1/M})$. (See Section \ref{2-3-0-section} for details.) Taking the limit   ${M\rightarrow \infty}$ we obtain the following list of Bethe permanent vectors based on all possible subsets $\beta\subset [4]$  of size $m{+}1 = 3$: $(1, 1, 1, 0)$, $(1, 1, 0, 1)$, $(1, 0, 1, 1)$, $(0,
  1, 1, 1)$. It can be easily checked that all the above vectors  satisfy the inequalities of the fundamental cone above, so they are all pseudocodewords. They give an upper bound on the minimum pseudo-weight of $8/3$, obtained for $M=1$ case in which the set of  perm-pseudocodewords listed in  Example \ref{example:1} is equal to the set of the degree-$M$ Bethe  permanent vectors described above. 
\end{example}

We revisit  Example \ref{example-motivation} in order to show the list of the Bethe permanent vectors for the matrix $\matr{H}$. 

\begin{example} \label{example-motivation-cont}
Let $M=3$ and $\matr{H} $ and $\theta$ like in Example \ref{example-motivation}.
We computed some of the length 12 Bethe permanent vectors and their pseudo-weights based on five column sets of length $3M+1=10$ and included them in Table \ref{table}, together with the perm-vectors based on the same sets of columns and their pseudo-weight for comparison. We used the Matlab program given in \cite{Vontobel:11:3:subm}.
Note that the size of the matrix is small in this example and so the permanent can be easily computed. If the matrix is large, in the order of $10^3$, the perm-vectors can not be computed anymore and the approximation given by the Bethe permanents will be valuable. 

\begin{table*}\begin{center}\begin{tabular}{|cccccccccccc||c|}
\hline
    2.3704 & 1.6875 & 1.6875 & 1.1250 & 1.1250 & 1.1250 & 1.1250 & 1.1250 & 1.1250 & 1.0000 &      0 &      0&8.8947\\\hline
    6  & 4 &  4 &  2 &  2 &  2 &  2 &  2 &  2 &  2 &  0 &  0 &8.1667\\\hline\hline
    1.6875 & 2.3704 & 1.6875 & 1.1250 & 1.1250 & 1.1250 & 1.1250 & 1.1250 & 1.1250 &      0 & 1.0000 &      0& 8.8947\\\hline4 &  6 &  4 &  2 &  2 &  2 &  2 &  2 &  2 &  0  & 2 &  0&8.1667 \\\hline\hline
    1.6875 & 1.6875 & 2.3704 & 1.1250 & 1.1250 & 1.1250 & 1.1250 & 1.1250 & 1.1250 &    0 &    0 & 1.0000&8.8947 \\\hline4 & 4 & 6 & 2 & 2 & 2 & 2 & 2 & 2 & 0 & 0 & 2& 8.1667 \\\hline\hline
 1.6875 & 1.6875 & 1.6875 & 0.0000 & 1.1250 & 1.1250 & 1.1250 & 1.1250     &    0 & 1.1250 &      0 & 1.1250&8.0000 \\ \hline4 & 4 & 4 & 0 & 2 & 2 & 2 & 2 & 0 & 2 & 0 & 2&8.0000 \\\hline\hline
    1.6875 & 1.6875 & 1.6875 & 1.1250 & 0.0000 & 1.1250 & 1.1250 & 1.1250    &     0 &      0 & 1.1250 & 1.1250&8.0000 \\\hline 4 & 4 & 4 & 2 & 0 & 2 & 2 & 2 & 0 & 0 & 2 & 2& 8.0000 \\\hline
    \end{tabular}\end{center}\caption{Pairs of Bethe permanent vectors and perm-vectors based on five sets $\beta$, together with their respective AWGNC pseudo-weights.}\label{table}\end{table*}\end{example}

\section{Equivalences of  the Conjecture and the Proof for Special Cases} 
\label{sec:results}
\subsection{An Equivalent Form of the Conjecture}\label{equiv:conj}
In the following we show the equivalence between Vontobel's conjecture for a general matrix to Vontobel's conjecture for a  matrix with a particular structure  (having a column of  weight 1)  for which only one of the inequalities in   \eqref{eq:fund:cone:def:2} needs to be verified. From this, we will show another equivalent description involving only square matrices.   The Bethe-perm vectors are based on a set $\beta$ of size $m+1$, therefore,  we will assume, without loss of generality,  that $n=m+1$. 
\begin{theorem} \label{theorem:reduction} The following statements are equivalent. \\
1) The conjecture holds for all $m\times (m+1)$  matrices $\matr{H}.$ \\
2) The conjecture holds for all $m\times (m+1)$ matrices $\matr{H}$  with a column of Hamming weight 1.\\
3) For any  $m\times (m+1)$ binary matrix $\matr{H}$ with the first column  equal to $[1~0~\cdots~ 0]^\tr$, the following inequality holds
\begin{align}\vomega_{\mathrm{B},1} \leq \sum_{l \in   \supp(\vect{R}_1)\setminus 1}\vomega_{\mathrm{B},l}.\label{ineq:new}\end{align} \\
4) For any  $m\times m$ binary (square) matrix $\matr{T}=(t_{ij})_{1\leq i,j\leq m}$, its Bethe permanent is less than or equal to its ``permanent-co(perm)factor expansion"  along any one of its rows,\footnote{By ``permanent-co(perm)factor expansion" we mean that the terms in the expansion sums are all taken with the positive sign and the Bethe permanent replaces the determinant in the definition of the cofactor.} i.e.,
\begin{align}\label{square:matrix:conj}\perm_{\mathrm{B}}\big( \matr{T}\big) \leq  \sum_{l\in[m]}
           t_{il}
           \cdot
           \perm_{\mathrm{B}}\big( \matr{T}_{[m]\setminus i, [m]\setminus  l} \big),  \end{align} for all $ i\in [m]$.
\end{theorem}

\begin{IEEEproof} 1) $\Rightarrow $ 2)  and 2) $\Rightarrow$  3) are  obvious (the inequality \eqref{ineq:new} is one of the inequalities in \eqref{eq:fund:cone:def:2}, which are satisfied  if we assume either 1) or 2) to be true. 
We prove 3) $\Rightarrow$  1), so we assume that the conjecture was proved for matrices with one column of weight 1, equal to $[1~0~\cdots~ 0]^\tr$ . We will show that we can now prove the conjecture for any other case.  

Let $\matr{H}$ be an  $m\times (m+1) $ binary matrix and  $\vomega_{\mathrm{B}}=(\vomega_{\mathrm{B},i})_{1\leq i\leq  m+1}$  be the Bethe-perm vector based on $[m+1]$. The matrix $\matr{H}$ and all its covers contain non-negative entries only (binary entries), therefore, the Bethe permanent of its submatrices are all non-negative giving that $\vomega_{\mathrm{B}}$ satisfies the inequalities \eqref{eq:fund:cone:def:1} of the fundamental cone. We will now look at the inequalities of the second type described in \eqref{eq:fund:cone:def:2}. The first row $\vect{R}_1$ contains $ \supp(\vect{R}_1)\geq 1$ ones (for zero support, it is trivial). Without  loss of generality, we can assume that the first nonzero entry of the first row is in the first column (we can always permute the columns of $\matr{H}$ which does not affect the value of the permanents and Bethe permanents, just their  order in the vector $\vomega_{\mathrm{B}}$.)
We need to show the inequality 
\begin{align*}\vomega_{\mathrm{B},1}\leq  \sum_{l \in   \supp(\vect{R}_1)\setminus 1}\vomega_{\mathrm{B},l} ,\end{align*} or equivalently,  
\begin{align*}\perm_{\mathrm{B}}\big( \matr{H}_{\beta \setminus 1} \big) \leq \sum_{l \in   \supp(\vect{R}_1)\setminus 1}\perm_{\mathrm{B}}\big( \matr{H}_{\beta \setminus l}\big).\end{align*} 

Let $\matr{\tilde H}$ be an $m\times (m+1)$ matrix with the first column equal to the length $m$ vector $[1~0~\cdots~ 0]^\tr$ and all the other columns equal to the corresponding columns of $\matr{H}$, therefore $\matr{\tilde H}$ is identical with $\matr{H}$ except for its first column which has weight 1, with the nonzero entry in the first position. Therefore $\perm_{\mathrm{B}}\big( \matr{\tilde H}_{\beta \setminus 1}\big) = \perm_{\mathrm{B}}\big( \matr{H}_{\beta \setminus 1}\big) $  (identical matrices) and,  for each $l\in   \supp(\vect{R}_1)\setminus 1$, $\perm_{\mathrm{B}}\big( \matr{\tilde H}_{\beta \setminus l}\big) \leq \perm_{\mathrm{B}}\big( \matr{H}_{\beta \setminus l}\big) $ (the two matrices are identical except for the first column  which is $[1~0~\cdots~ 0]^\tr$ for $\matr{\tilde H}_{\beta \setminus l}$ and $[1~*~\cdots~ *]$ for the matrix $\matr{H}_{\beta \setminus l}$, where $*$ stands for an entry of 1 or 0). 
 
The matrix $\matr{\tilde H}$ is in the form assumed in 3)  for which we assumed inequality \eqref{ineq:new} to hold. 
We obtain 
\begin{align*}\perm_{\mathrm{B}}\big( \matr{H}_{\beta \setminus 1} \big)=&\perm_{\mathrm{B}}\big( \matr{\tilde H}_{\beta \setminus 1} \big)\\ \leq& \sum_{l \in   \supp(\vect{R}_1)\setminus 1}\perm_{\mathrm{B}}\big( \matr{\tilde H}_{\beta \setminus l}\big)  \\\leq& \sum_{l \in   \supp(\vect{R}_1)\setminus 1}\perm_{\mathrm{B}}\big( \matr{H}_{\beta \setminus l}\big) \end{align*} 
 which was the desired inequality. 
 
 We proceed similarly for all the other inequalities corresponding to the first row $$\vomega_{\mathrm{B},l}
      \leq
          \sum_{l' \in \supp(\vect{R}_1) \setminus l} \!\!
          \vomega_{\mathrm{B},l'} $$
   for all  
                $l \in \supp(\vect{R}_1)\setminus 1$ and reduce the problem to the case for which the $l$th column of $\matr{H}$ is the vector $[1~0~\cdots~ 0]^\tr$, for which we have assumed to have proved  the conjecture.  In order to show that the vector satisfies all the inequalities associated with any row $\vect{R}_i,$ $2\leq i\leq m$, we  first permute the rows of $\matr{H}$ to place $\vect{R}_i$ in the first position (this does not alter the vector  $\vomega_{\mathrm{B}}$) and then use the same reasoning above to show that all the inequalities  in \eqref{eq:fund:cone:def:2} associated to $\vect{R}_i$ are satisfied. This concludes the proof of 3) $ \Rightarrow$ 1). 
               
       The equivalence 3) $\Longleftrightarrow$ 4) is immediate once we observe that any square  matrix  $\matr{T}=(t_{ij})_{1\leq i,j\leq m}$ can be seen as a submatrix $\matr{H}_{\beta \setminus 1} $ of an $m\times (m+1)$ matrix $\matr{H}$ with its first column equal to $[1~0~\cdots~ 0]^\tr$. 
       Then, we have the two equalities  \begin{align*}\perm_{\mathrm{B}}\big(\matr{T}\big)&=\perm_{\mathrm{B}}\big( \matr{H}_{\beta \setminus 1} \big) =\vomega_{\mathrm{B},1} \end{align*}
       and \begin{align*} \sum_{l\in[m]}
           t_{1l}
           \cdot
           \perm_{\mathrm{B}}\big( \matr{T}_{[m]\setminus 1, [m]\setminus  l} \big) &= \\ \sum_{l \in   \supp(\vect{R}_1)\setminus 1}\perm_{\mathrm{B}}\big( \matr{H}_{\beta \setminus l}\big)&= \sum_{l \in   \supp(\vect{R}_1)\setminus 1}\vomega_{\mathrm{B},l},\end{align*} 
   due to the fact that   $\perm_{\mathrm{B}}\big( \matr{H}_{\beta \setminus l}\big)=\perm_{\mathrm{B}}\big( \matr{T}_{[m]\setminus 1, [m]\setminus  l} \big)$ from co(perm)factor-expanding the permanent along the first column of weight 1. 
  Therefore, we can easily see now the equivalence of the two inequalities 
  \begin{align*}\big\{\vomega_{\mathrm{B},1} & \leq \sum_{l \in   \supp(\vect{R}_1)\setminus 1}\vomega_{\mathrm{B},l}\big\} \Longleftrightarrow  \\  
\big\{\perm_{\mathrm{B}}\big( \matr{T}\big) & \leq  \sum_{l\in[m]}
           t_{1l}
           \cdot
           \perm_{\mathrm{B}}\big( \matr{T}_{[m]\setminus 1, [m]\setminus  l} \big)\big\}.\end{align*}
       \end{IEEEproof}
\begin{remark}
Therefore, in order to prove the conjecture, we can assume that  $\matr{H}$ has its first  column equal to  $[1~0~\cdots~ 0]^\tr$ and prove that: 
\begin{align*}\vomega_{\mathrm{B},1} & \leq \sum_{l \in   \supp(\vect{R}_1)\setminus 1}\vomega_{\mathrm{B},l}.\end{align*}
Note that in this case, $\vomega_{\mathrm{B},1}\geq \vomega_{\mathrm{B},l}$ for all  $l \in   \supp(\vect{R}_1)\setminus 1$, so the first component is the largest among the components indexed by the $\supp(\vect{R}_1)$. 
 \end{remark}
 In addition, in most our considerations, we will show that $\vomega_{\mathrm{B},M}$ is a pseudocodeword,  for all $M\geq 1$. Then, by taking the limit it will follow that $\vomega_{\mathrm{B}}\in \fch{K}{H}$.  We state and prove this briefly below.

 \begin{lemma}\label{one:only}
 Let $\code{C}$ be a  binary linear code described by a
  parity-check matrix $\matr{H}\in \Ftwo^{m \times n}$, $m < n$,  and $\beta$ be a
  size-$(m{+}1)$ subset $\beta$ of $[n]$. Let $\vomega_{\mathrm{B},M} $ and $\vomega_{\mathrm{B}}$ based on $\beta$ defined as in Definition \ref{def:perm:bethe:vector:1}.  If $\vomega_{\mathrm{B},M}\in \fch{K}{H}$, for all integers $M\geq 1$, then $\vomega_{\mathrm{B}}\in \fch{K}{H}$. \end{lemma}
\begin{IEEEproof}  Since $\vomega_{\mathrm{B},M}\in \fch{K}{H}$, for all $M\geq 1$, each of the inequalities in  \eqref{eq:fund:cone:def:1} and  \eqref{eq:fund:cone:def:2} is satisfied by $\vomega_{\mathrm{B},M, i}$, $i\in [n]$. Taking the limit when $M\rightarrow \infty$ gives that $\vomega_{\mathrm{B},i}$  must also satisfy the same inequalities. It follows that $ \vomega_{\mathrm{B}}\in \fch{K}{H}$.
\end{IEEEproof}

\subsection{Case of Row Degrees $\leq 2$} \label{sec:case:degree:2}

If a row in the matrix $\matr{H}$ has weight 2 or lower, then the inequalities \eqref{eq:fund:cone:def:2} associated with it are always satisfied. It follows that, if the matrix has all rows of degree $\leq 2$, the Bethe permanent vector is a  pseudocodeword. 
\begin{lemma} \label{degree2} Let $\matr{H}$ be an $m\times (m+1) $ binary matrix with the first row of degree 2 or lower. Then, the Bethe permanent vector $ \vomega_{\mathrm{B}}$ satisfies the 
inequalities \eqref{eq:fund:cone:def:2} associated with the first row.
\end{lemma}

\begin{IEEEproof}
Let $\matr{H}$ of the form
$\matr{H}=\begin{bmatrix} 1&1 &{0}\\ \vect{c}_1&\vect{c}_2 &\matr{A}\end{bmatrix}$, where  $\vect{c}_i\in \Ftwo^{(M-1)\times 1}$, $\matr{A}\in \Ftwo^{(M-1)\times (M-1)}$ and the all-zero matrix $0$ of size $1 \times (M-1)$. The two components $ \vomega_{\mathrm{B}, 1}$ and $ \vomega_{\mathrm{B}, 2}$ are
\begin{align*}& \vomega_{\mathrm{B}, 1}=\perm_{\mathrm{B}}\big( \begin{bmatrix} 1 &{0}\\ \vect{c}_2 &\matr{A}\end{bmatrix} \big)= \perm_{\mathrm{B}}(\matr{A}) \\ & \vomega_{\mathrm{B}, 2}=\perm_{\mathrm{B}}\big( \begin{bmatrix} 1 &{0}\\ \vect{c}_1 &\matr{A}\end{bmatrix} \big)= \perm_{\mathrm{B}}(\matr{A}).  \end{align*}
Therefore \begin{align*} &\vomega_{\mathrm{B}, 1} = \vomega_{\mathrm{B}, 2} \Longleftrightarrow   \big \{\vomega_{\mathrm{B}, 1} \leq  \vomega_{\mathrm{B}, 2} {\rm ~and~} \vomega_{\mathrm{B}, 1} \geq  \vomega_{\mathrm{B}, 2}\big\}, \end{align*} yielding that $ \vomega_{\mathrm{B}}$ is a pseudocodeword.

 The cases for which the row is all zero or has weight 1 are trivial, since $ \vomega_{\mathrm{B}}=\zero$ in those cases.
\end{IEEEproof}

\begin{corollary}
Let $\matr{H}$ be an $m\times (m+1) $ binary matrix that has  all its rows of degree 2 or lower. Then the Bethe permanent vector $ \vomega_{\mathrm{B}}$ is a pseudocodeword.
\end{corollary}
\begin{IEEEproof}
We apply Lemma \ref{degree2} for each row of degree 2 to obtain the claim. The inequalities associated with the rows of degree 1 are satisfied trivially since the component equal  to the Bethe permanent of the submatrix obtained by erasing the column indexed by the nonzero entry is zero. This is due to the all-zero row that the submatrix was left with after removing the column with the only nonzero entry on that row.      
\end{IEEEproof}
\subsection{Case  of $\matr{H}$ of the Form \eqref{matrix-particular-star}}\label{case:conj}

In   Examples \ref{example:11} and \ref{example-motivation-cont},  we observed that the sets of the degree-$M$ Bethe  permanent vectors and the Bethe permanent vectors based on all possible subsets $\beta$ form two sets of pseudocodewords. In fact it is enough to observe  that the degree-$M$ Bethe  permanent vectors  based on all possible subsets $\beta$ are all pseudocodewords to conclude, by taking the limit, that the same  is true for the given matrix,  confirming thus the conjecture for this particular matrix
  $\matr{H}$.  In this section, we show that this stronger version of the conjecture is always true for a more general case, that of  
\begin{align}\label{matrix-particular-star}
\matr{H}=\left(\begin{matrix}1&1&1&\cdots&1\\ *&1&1&\cdots &1\\ \vdots&\vdots&\vdots&&\vdots\\
*&1&1&\cdots&1\end{matrix}\right)\in \Ftwo^{m\times (m+1)},\end{align} 
where $m\geq 2$ is an integer,  and $*$ can be  either 0 or 1, 
and compute exactly the degree-$M$ Bethe  permanent vectors and the Bethe permanent vectors  for the case $m=2$.
Similar to our reasoning in Section \ref{equiv:conj}, we will show briefly that  it is enough to show  the conjecture for a matrix $\matr{H}$ of  the form 
\begin{align}\label{matrix-particular}
\matr{H}=\left(\begin{matrix}1&1&1&\cdots&1\\ 0&1&1&\cdots &1\\ \vdots&\vdots&\vdots&&\vdots\\
0&1&1&\cdots&1\end{matrix}\right)\in \Ftwo^{m\times (m+1)}. \end{align}  Equivalently, we only need to show 
  the inequality \eqref{ineq:new} $\vomega_{\mathrm{B},1}  \leq \sum_{l \in   [m+1]\setminus 1}\vomega_{\mathrm{B},l}.$ 
Indeed, the inequalities \eqref{eq:fund:cone:def:2} associated with any of the $i$th row of $\matr{H}$  corresponding to a nonzero entry in its  first column are  equivalent to the inequality  \eqref{ineq:new}  since 
$\vomega_{\mathrm{B},1}\geq \vomega_{\mathrm{B},l}$ for all  $l \in   [m+1]\setminus 1$. The inequalities associated with any of the $i$th row of $\matr{H}$  corresponding to a zero  entry are trivially satisfied since all components $\vomega_{\mathrm{B},l} $ are equal, for all $ 2\leq l\leq m+1$. 

For the remainder  of this section, we will assume $\matr{H}$ to be of the form \eqref{matrix-particular}. Although this matrix looks simple, the proof requires some thought and manipulations.

\begin{theorem}\label{thm:conjecture}
Let  $\matr{H}$ be of the form \eqref{matrix-particular}. 
Then, for all $M\geq 1$, its degree-$M$ Bethe permanent vectors $\vomega_{\mathrm{B},M}$ and its Bethe permanent vector $\vomega_{\mathrm{B}}$ based on $\beta\defeq [m+1]$ are pseudocodewords.  
\end{theorem}
We will call these pseudocodewords the {\em degree-$M$ Bethe permanent pseudocodeword  based on $\beta$} and the  {\em Bethe permanent pseudocodeword based on $\beta$}, respectively, and, when we are considering both sets, we will call them the {\em Bethe pseudocodewords}. 

As we mentioned already in Lemma \ref{one:only}, in proving Theorem \ref{thm:conjecture}, it is enough to show that $\vomega_{\mathrm{B},M}\in \fch{K}{H}$, for all $M\geq 1$. 

Before proving this theorem, we need a few lemmas and theorems. 

Let $\theta=\vect{1}_{m\times m}$ be the all-one matrix of size $m\times m$.  
Let $I$ denote the identity matrix of size $M\times M$
and for all $m\geq 1$, let 
 \begin{align*}
&p_{m-1,m, M}\defeq \\&\sum\limits_{\matr{Q}\in \Psi_{m-1,m, M}}\!\!\!\!\!\!\!\!\perm\begin{bmatrix} I &I&\cdots&I\\ Q_{11}& Q_{12}&\cdots &Q_{1,m}\\ \vdots &\vdots&&\vdots\\ Q_{m-1,1}&Q_{m-1,2}&\cdots&Q_{m-1,m} \end{bmatrix},  \\ &q_{m-1,m,M}\defeq \frac{p_{m-1,m,M}}{(M!)^{(m-1)m}}=\frac{p_{m-1,m,M}}{\left|\Psi_{m-1,m,M}\right|},\end{align*}
\begin{align*}
&p_{m,M}\defeq \\&\sum\limits_{P\in \Psi_{m-1,M}}\!\!\!\!\perm\begin{bmatrix} I &I&\cdots&I\\ I& P_{11}&\cdots &P_{1,m-1}\\ \vdots &\vdots&&\vdots\\ I&P_{m-1,1}&\cdots&P_{m-1,m-1} \end{bmatrix}{~\rm and~}\\  &q_{m,M}\defeq \frac{p_{m,M}}{(M!)^{(m-1)^2}}=\frac{p_{m,M}}{\left|\Psi_{m,M}\right|},
\end{align*}
\begin{lemma} \label{recursive} We have the following relations between $q_{m,M}$, $q_{m-1,m,M}$, and the average $ {\big<\perm(\theta^{\uparrow\matr{R}})\big>_{\matr{R}\in \Psi_{m,M}}}$:
$$q_{m,M}=q_{m-1,m,M}={\big<\perm(\theta^{\uparrow\matr{R}})\big>_{\matr{R}\in \Psi_{m,M}}}.$$
\end{lemma}
\begin{IEEEproof}
 Any $\matr{R}$-lifting of the matrix $\theta$, $\matr{R}=(R_{ij})\in\Psi_{m,M}$, can be written as
\begin{align*} \label{matrix:general1}
\theta ^{\uparrow\matr{R}}&=\begin{bmatrix} 
R_{11}& R_{12}&\cdots & R_{1m} \\
R_{21}&R_{22} &\cdots& R_{2m}\\ 
\vdots&\vdots&\cdots&\vdots\\
R_{m1}&R_{m2} &\cdots& R_{mm}
\end{bmatrix}.\end{align*} 
Because the permanents are not affected by row or column permutations, 
we can apply permutations of columns  to reduce the lifting matrix to the simpler form
\begin{align} 
&\begin{bmatrix} 
I          &I                                                                              &\cdots  & I \\
R_{21}R_{11}^\tr         & R_{22} R_{12}^\tr            &\cdots  & R_{2m}R_{1m}^\tr\\ 
\vdots&\vdots                                                                       &\cdots  &\vdots\\
R_{m1}R_{11}^\tr         & R_{m2} R_{12}^\tr &\cdots  &  R_{mm}R_{1m}^\tr
\end{bmatrix}\nonumber \\\defeq& \begin{bmatrix} 
I &I &\cdots & I \\
U_{11}&U_{12} &\cdots& U_{1m}\\ 
\vdots&\vdots&\cdots&\vdots\\
U_{m-1,1}&U_{m-1,2} &\cdots& U_{m-1,m}
\end{bmatrix},\end{align} 
followed by row permutations to reduce it to 
\begin{align} 
&\begin{bmatrix} 
I          &I                                                                              &\cdots  & I \\
I         &U_{11}^\tr U_{12}            &\cdots  & U_{11}^\tr U_{1m}\\ 
\vdots&\vdots                                                                       &\cdots  &\vdots\\
I         & U_{m-1,1}^\tr U_{m-1,2} &\cdots  & U_{m-1,1}^\tr U_{m-1,m}
\end{bmatrix}\nonumber \\\defeq&\label{standard:matrix}\begin{bmatrix} 
I &I &\cdots & I \\
I&Q_{11} &\cdots& Q_{1,m-1}\\ 
\vdots&\vdots&\cdots&\vdots\\
I&Q_{m-1,1} &\cdots& Q_{m-1,m-1}
\end{bmatrix},\end{align} 
where $\matr{U}=(U_{ij})\in \Psi_{m-1,m,M}$ and  $\matr{Q}=(Q_{ij})\in \Psi_{m-1,M}$. Therefore, the average ${\big<\perm(\theta^{\uparrow\matr{R}})\big>_{\matr{R}\in \Psi_{m,M}}}$ of the permanents of $\theta ^{\uparrow\matr{R}} $ over all matrices $\matr{R}\in \Psi_{m,M}$ equals both $q_{m,M}$ and $q_{m-1,m,M}$. 
\end{IEEEproof}

With these notations, the inequality \eqref{ineq:new} needed to show the conjecture for the matrix $\matr{H}$ of the form \eqref{matrix-particular} is equivalent to  \begin{center} \fbox{$q_{m,M}\leq m^M \cdot q_{m-1,M}$}\end{center}
\begin{lemma} \label{recursive1} Let $m\geq 2$ and $M\geq 1$. Let $\beta_i \subseteq [M]$, $1\leqslant i \leqslant m$, such that    $\{ [M]\setminus \beta_1, \ldots, [M]\setminus\beta_m\}$ is a partition of $[M]$ (i.e.,  $m$ disjoint sets of column indices with union equal to $[M]$). Let 
$\matr{Q}=(Q_{ij})\in \Psi_{m-1,m, M}$. 
Then \begin{align}\label{recursive-ineq1}
&\sum\limits_{\tiny\begin{matrix}Q_{ij}\in \mathcal{P}_M\\ i\in[m-1]\\ j\in [m]\end{matrix}}\!\!\!\!\perm\begin{bmatrix}  Q_{11,\beta_1}& Q_{12,\beta_2}&\cdots &Q_{1,m, \beta_m}\\ \vdots&\vdots&&\vdots\\ Q_{m-1,1,\beta_1}&Q_{m-1,2,\beta_2}&\cdots &Q_{m-1,m, \beta_m}\end{bmatrix} \nonumber \\&\leq\sum\limits_{\tiny\begin{matrix}P_{ij}\in \mathcal{P}_M\\ i,j\in [m-1]\end{matrix}}\perm\begin{bmatrix} P_{11}&\cdots &P_{1,m-1}\\ \vdots&&\vdots\\ P_{m-1,1}&\cdots &P_{m-1,m-1}\end{bmatrix}. \end{align}
\end{lemma}
\begin{IEEEproof} See Appendix A.
\end{IEEEproof}
\begin{corollary} \label{recursive2} Let $q_{m,M}={\big<\perm(\theta^{\uparrow\matr{R}})\big>_{\matr{R}\in \Psi_{m,M}}}$, for $m\geq 1$ (see Lemma \ref{recursive}). Then
 \begin{align} \label{recursive-ineq2}  q_{m,M}\leq m^M q_{m-1,M}.\end{align}
\end{corollary} 
\begin{IEEEproof}
The permanent of a matrix $\matr{Q}\in\Psi_{m, M}$ as in \eqref{matrix:general1} (i.e., such that the first block row has entries only identity matrices) is computed by summing all products of entries such that each row and each column contribute to the product exactly once. The first $M$ rows can contribute with 1s from the identity matrices on the top block row by taking the 1 entries from the columns indexed by   
$[M]\setminus \beta_1$ of the first block of $M$ columns,  $[M]\setminus \beta_2$ of the second block of $M$ columns, such that  $[M]\setminus \beta_2\subseteq [M]\setminus \left([M]\setminus \beta_1\right)=\beta_1$, and finishing with $[M]\setminus \beta_m$ of the $m$th block of $M$ columns.  In any such choice of $M$ entries of 1 from the first $M$ rows, $[M]\setminus \beta_1, [M]\setminus \beta_2, \ldots, [M]\setminus \beta_m$ must be a partition of $[M]$.  Let  $\left|\beta_i\right|=r_i$, then $r_1+\cdots+r_m=(m-1)M$. Note that the columns in the sets $[M]\setminus \beta_1, [M]\setminus \beta_2, \ldots, [M]\setminus \beta_m$ cannot contribute anymore to the rest of the product entries. There are $ {M \choose {M-|\beta_1|}}={M \choose {M-r_1}}={M \choose r_1}$ ways to choose the $r_1$ columns from the first block of $M$ columns, $0\leq r_1\leq M$. Then there are $ {r_1 \choose r_2}$ ways to choose the $r_2$ columns  from the second  block of $M$ columns indexed by the  remaining $r_1=|\beta_1|$ indices,  $0\leq r_2\leq r_1$, etc., finishing with  ${r_{m-2} \choose r_{m-1}}$ ways to choose the $r_{m-1}$ columns in the $(m-1)$th block of $M$ columns,  $0\leq r_{m-1}\leq r_{m-2}$. The last $M$ columns  are uniquely determined by the fact that the corresponding  rows in the top block should give a partition of $[M]$. Therefore, we can choose the 1s in the first $M$ rows in exactly   $\sum\limits_{r_1=0}^{M}{M \choose r_1} \sum\limits_{r_2=0}^{r_1} {r_1 \choose r_2}\ldots \sum\limits_{r_{m-1}=0}^{r_{m-2}}{r_{m-2} \choose r_{m-1}}$ ways, followed by  products with entries in the next $M+1, M+2, \ldots, (m-1)M+1, (m-1)M+2, \ldots, mM$ rows.  Equivalently, we can compute and upper bound $q_{m,M}$ in the following  way. 
  \begin{align*}
&q_{m,M}= q_{m-1,m,M}=\frac{\sum\limits_{r_1=0}^{M} {M \choose r_1} \sum\limits_{r_2=0}^{r_1} {r_1 \choose r_2}\ldots \sum\limits_{r_{m-1}=0}^{r_{m-2}}{r_{m-2} \choose r_{m-1}}}{(M!)^{(m-1)m}}\times \\&\times\!\!\! \sum\limits_{\tiny\begin{matrix}Q_{ij}\in \mathcal{P}_M\\ i\in[m-1]\\ j\in [m]\end{matrix}}\!\!\!\!\!\perm\!\!\begin{bmatrix}  Q_{11,\beta_1}& Q_{12,\beta_2}&\cdots &Q_{1,m, \beta_m}\\ \vdots&\vdots&&\vdots\\ Q_{m-1,1,\beta_1}&Q_{m-1,2,\beta_2}&\cdots &Q_{m-1,m, \beta_m}\end{bmatrix} \\ 
&\leq \sum\limits_{r_1=0}^{M} {M \choose r_1} \sum\limits_{r_2=0}^{r_1} {r_1 \choose r_2}\ldots \sum\limits_{r_{m-1}=0}^{r_{m-2}}{r_{m-2} \choose r_{m-1}} q_{m-1,M} \\&=q_{m-1,M} \sum\limits_{r_1=0}^{M} {M \choose r_1} \sum\limits_{r_2=0}^{r_1} {r_1 \choose r_2}\ldots \sum\limits_{r_{m-1}=0}^{r_{m-2}}{r_{m-2} \choose r_{m-1}}\\&= m^M q_{m-1,M}. \end{align*}
The inequality above holds due to Lemma \ref{recursive1} and the fact that $\frac{1}{(M!)^{(m-1)m}}\leq \frac{1}{(M!)^{(m-1)^2}}$.  \end{IEEEproof} 

\vspace{.3in}

\begin{IEEEproof}[{Proof of Theorem \ref {thm:conjecture}}] We need to show that the inequalities  \eqref{eq:fund:cone:def:1} and  \eqref{eq:fund:cone:def:2} of the fundamental cone described in Definition \ref{def:pseudocodewords} hold for  $\vomega_{\mathrm{B},M}$. The inequality  \eqref{eq:fund:cone:def:1} is trivially satisfied since the matrix has non-negative  entries; thus the permanent is a sum of non-negative entries, therefore non-negative, giving that each component is non-negative.  In order to check inequality \eqref{eq:fund:cone:def:2},  we need to show that  each  component we choose from  $\vomega_{\mathrm{B},M}$ is less than or equal to the sum of the remaining components.  From Lemma \ref{recursive},  we know that 
$\vomega_{\mathrm{B},M}=(q_{m,M}^{1/M}, q_{m-1,M}^{1/M},\ldots, q_{m-1,M}^{1/M}).  $ If the chosen component is $q_{m,M}^{1/M} $, then Cor. \ref{recursive2} gives us the  needed inequality  $q_{m,M}^{1/M}\leq m q_{m-1,M}^{1/M}$.  
If the component is equal to $q_{m-1,M}^{1/M}$, then among the remaining components there is at least one component equal to $q_{m-1,M}^{1/M}$, which leads to the needed fundamental cone inequality being satisfied. Therefore, $\vomega_{\mathrm{B},M}$ is a pseudocodeword.
Taking the limit and applying Lemma \ref{one:only} we get that $\vomega_{\mathrm{B}}$ is also a pseudocodeword. 
 \end{IEEEproof}

\subsection{Case  of $\matr{H}$ of the Form \eqref{matrix-particular-case}}\label{case:conj2}

 In this section, we will show how to adjust the proof in the previous section to show the conjecture  for a matrix in slightly more general form than that in  \eqref{matrix-particular-star}: we will assume that  the first row of $\matr{H}$ contains less than $m+1$ ones
\begin{align}\label{matrix-particular-case}
\matr{H}=\left(\begin{matrix}1&1&1&\cdots&1&0&\cdots &0\\ 0&1&1&\cdots &1&1&\cdots &1 \\ \vdots&\vdots&\vdots&&\vdots&\vdots &&\vdots\\
0&1&1&\cdots&1&1&\cdots &1\end{matrix}\right)\in \Ftwo^{m\times (m+1)}. \end{align} For simplicity, we will assume only one extra zero on the first row and prove for  the simpler looking matrix 
\begin{align}\label{matrix-particular-case-one}
\matr{H}=\left(\begin{matrix}1&1&1&\cdots&1&0\\ 0&1&1&\cdots &1&1 \\ \vdots&\vdots&\vdots&&\vdots&\vdots\\
0&1&1&\cdots&1&1\end{matrix}\right)\in \Ftwo^{m\times (m+1)} \end{align}  
the inequality  \begin{center}\fbox{$t_{m,M} \leq (m-1)^M\cdot q_{m-1,M},$}\end{center}
where 
\begin{align*}
&t_{m,M}\defeq\\ &\frac{\sum\limits_{Q\in \Psi_{m-1, M}}\!\!\!\!\perm\begin{bmatrix} I&\cdots&I&0\\  Q_{11}&\cdots &Q_{1,m-1}&I\\ \vdots&&\vdots&\vdots \\ Q_{m-1,1}&\cdots&Q_{m-1,m-1} &I\end{bmatrix}}{(M!)^{(m-1)^2}}.
\end{align*}
Note that for the matrix in  \eqref{matrix-particular-case-one}, $\vomega_{\mathrm{B},M}$ is equal to  $$\vomega_{\mathrm{B},M}=\left(t_{m,M}^{1/M}, q_{m-1,M}^{1/M}, q_{m-1,M}^{1/M},\cdots,   q_{m-1,M}^{1/M}\right).$$
Therefore, by adding one zero to the all-one matrix $\theta =\matr{H}_{[m+1] \setminus 1}$, with $\matr{H}$ given  in \eqref{matrix-particular}, we decrease the upper bound on its degree-$M$ Bethe permanent from $m\cdot  q_{m-1,M}^{1/M}$ (see  \eqref{recursive-ineq2}) to $(m-1)\cdot  q_{m-1,M}^{1/M}$ for the submatrix of the matrix in \eqref{matrix-particular-case-one} with columns indexed by the same set ${[m+1] \setminus 1}$.
Similarly, when we further increase the number of zeros on the first row of the all-one matrix,  we further decrease the upper bound on its degree-$M$ Bethe permanent from $m\cdot  q_{m-1,M}^{1/M}$ to $(\supp(\vect{R}_1)-1)\cdot  q_{m-1,M}^{1/M}$.  Therefore, any extra zero on one row of the all-one matrix, produces a decrease in the upper bound of its degree-$M$ Bethe permanent  by one quantity $q_{m-1,M}^{1/M}$. 



For the remainder  of this section, we will assume $\matr{H}$ to be in the form \eqref{matrix-particular-case-one}. 

\begin{theorem}\label{thm:conjecture:1}
Let  $\matr{H}$ be  of the form in \eqref{matrix-particular-case-one}. 
Then, for all $M\geq 1$, its degree-$M$ Bethe permanent vectors $\vomega_{\mathrm{B},M}$ and its Bethe permanent vector $\vomega_{\mathrm{B}}$ based on $\beta\defeq [m+1]$ are pseudocodewords for $\matr{H}$.  
\end{theorem}
\begin{IEEEproof} The proof follows the same reasoning and steps of  the proofs of Theorem \ref{thm:conjecture} and  Corollary \ref{recursive2}.
\begin{align*}
&t_{m,M}=\frac{\sum\limits_{r_1=0}^{M} {M \choose r_1} \sum\limits_{r_2=0}^{r_1} {r_1 \choose r_2}\ldots \sum\limits_{r_{m-2}=0}^{r_{m-3}}{r_{m-3} \choose r_{m-2}}}{(M!)^{(m-1)^2}}\times \\&\times\!\!\!\!\!\! \sum\limits_{\tiny\begin{matrix}Q_{ij}\in \mathcal{P}_M\\ i,j\in[m-1]\end{matrix}}\!\!\!\!\!\perm\!\!\begin{bmatrix}  Q_{11,\beta_1}\!\!&\!\! Q_{12,\beta_2}\!\!&\!\!\cdots \!\!&\!\!Q_{1,m-1, \beta_{m-1}}\!\!\!&\!\!I\\ \vdots\!\!&\!\!\vdots\!\!&\!\!\!\!&\!\!\vdots\!\!\!&\!\!\vdots\\ Q_{m-1,1,\beta_1}\!\!&\!\!Q_{m-1,2,\beta_2}\!\!&\!\!\cdots \!\!&\!\!Q_{m-1,m-1, \beta_{m-1}}\!\!\!&\!\!I\end{bmatrix} \\ 
&\leq \sum\limits_{r_1=0}^{M} {M \choose r_1} \sum\limits_{r_2=0}^{r_1} {r_1 \choose r_2}\ldots \sum\limits_{r_{m-2}=0}^{r_{m-3}}{r_{m-3} \choose r_{m-2}} q_{m-1,M} \\&=q_{m-1,M} \sum\limits_{r_1=0}^{M} {M \choose r_1} \sum\limits_{r_2=0}^{r_1} {r_1 \choose r_2}\ldots \sum\limits_{r_{m-2}=0}^{r_{m-3}}{r_{m-3} \choose r_{m-2}}\\&= (m-1)^M q_{m-1,M}. \end{align*}
The inequality above can be obtained by modifying the proof of Lemma \ref{recursive1} to allow for a column of identity matrices to obtain  
\begin{align*}
&
\sum\limits_{\tiny\begin{matrix}Q_{ij}\in \mathcal{P}_M\\ i,j\in[m-1]\end{matrix}}\!\!\!\!\!\!\!\perm\!\!\begin{bmatrix}  Q_{11,\beta_1}\!\!&\!\! Q_{12,\beta_2}\!\!&\!\!\cdots \!\!&\!\!Q_{1,m-1, \beta_{m-1}}\!\!&\!\!I\\ \vdots\!\!&\!\!\vdots\!\!&\!\!\!\!&\!\!\vdots\!\!&\!\!\vdots\\ Q_{m-1,1,\beta_1}\!\!&\!\!Q_{m-1,2,\beta_2}\!\!&\!\!\cdots \!\!&\!\!Q_{m-1,m-1, \beta_{m-1}}\!\!&\!\!I\end{bmatrix}\\
&\leq \sum\limits_{\tiny\begin{matrix}P_{ij}\in \mathcal{P}_M\\ i\in [m-1]\\j\in [m-2]\end{matrix}}\perm\begin{bmatrix} P_{11}&\cdots &P_{1,m-2}&I\\ \vdots&&\vdots&\vdots\\ P_{m-1,1}&\cdots &P_{m-1,m-2}&I \end{bmatrix}.
\end{align*}
\end{IEEEproof}
\subsection{Case  of $\matr{H}$ of the Form \eqref{matrix-particular-star2}}\label{case:conj3}

The natural next step to consider is  that of a matrix $\matr{H}$ with allowed  zeros in the first two rows and ones elsewhere. We will consider the simplest case and discuss the problems that this case  presents. 
Let 
\begin{align}\label{matrix-particular-star2}
\matr{H}=\left(\begin{matrix}1&1&1&\cdots&1&0\\ 0&1&1&\cdots &0&1 \\ \vdots&\vdots&\vdots&&\vdots&\vdots\\
0&1&1&\cdots&1&1\end{matrix}\right)\in \Ftwo^{m\times (m+1)} \end{align}  
and 
 \begin{center}\fbox{${\hat t}_{m,M}^{1/M} \leq (m-1)\cdot t_{m-1,M}^{1/M} +q_{m-1,M}^{1/M},$} \label{asymm}\end{center}
where 
\begin{align*}
&{\hat t}_{m,M}\defeq\\ &\frac{\sum\limits_{Q\in \Psi_{m-1, M}}\!\!\!\!\perm\begin{bmatrix} I&\cdots&I&0\\  Q_{11}&\cdots &0&I\\ \vdots&&\vdots&\vdots \\ Q_{m-1,1}&\cdots&Q_{m-1,m-1} &I\end{bmatrix}}{(M!)^{(m-1)^2-1}}.
\end{align*}
Note that for the matrix in  \eqref{matrix-particular-case-one}, $\vomega_{\mathrm{B},M}$ is equal to $$\vomega_{\mathrm{B},M}=\left(\hat{t}_{m,M}^{1/M}, t_{m-1,M}^{1/M},\cdots,   t_{m-1,M}^{1/M}, q_{m-1,M}^{1/M},  t_{m-1,M}^{1/M}\right).$$

Modifying the proof of  Theorem \ref{thm:conjecture:1}  to allow for an extra zero is not easy. We will illustrate this in the following numerical example.
\begin{example} Let $\matr{H}=\left(\begin{matrix}1&1&1&0\\0&1&0&1\\0&1&1&1\end{matrix}\right)$. Then
$$\vomega_{\mathrm{B},M}=\left({\hat t}_{3,M}^{1/M}, 1, q_{2,M}, 1 \right)=\left({\hat t}_{3,M}^{1/M}, 1, (M+1)^{1/M}, 1 \right).$$
For the last equality, see Theorem \ref{2-3case}.  Hence, in order to show that $\vomega_{\mathrm{B},M}$ is a pseudocodeword,  we need to show that ${\hat t}_{3,M}^{1/M} \leq 1+(M+1)^{1/M}$. 
In Theorem \ref{3-3-0case}, we compute an exact formula for ${\hat t}_{3,M}^{1/M}$ and based on it, in Corollary \ref{3-4-0pseudocdewods}, we show the needed inequality. 
\end{example} 

The difficulty of this case lies in the fact that on the right hand side of the inequality  ${\hat t}_{m,M}^{1/M} \leq (m-1)\cdot t_{m-1,M}^{1/M} +q_{m-1,M}^{1/M},$ there are $M$th roots of non-equal terms. This case cannot be solved in the way we solved the two cases of Sections \ref{case:conj} and \ref{case:conj2}, a  different idea is needed. However, for matrices with a lot of zeros, we might be able to compute exactly the  degree-$M$ Bethe permanents using  combinatorial arguments, by 
showing that the permanents of a block $mM\times mM$ matrix is equal to the permanent of a smaller size matrix. For example, in Appendix \ref{exact:q2}, we show that  $$q_{2,M}={{\sum\limits_{P\in \mathcal{P}_M}\perm(I+P)}\over {M!}}$$ and in Appendix \ref{exact:t3}, that $${\hat t}_{3,M}= {{\sum\limits_{P,Q\in \mathcal{P}_M}\perm(I+P+Q)}\over {(M!)^2}}$$ and that  $${t}_{3,M}={{\sum\limits_{P,Q,R\in \mathcal{P}_M}\perm(I+P+Q+R)}\over {(M!)^3}}$$
The expressions on the right  can be easily computed for any number of matrices in the sum and using  Maple, the needed inequality can be in many cases  verified. The following example illustrates this idea. 
 \begin{example}
 Let $$\matr{H}\defeq\begin{bmatrix} 1&1&1&0&0\\0&1&0&1&0\\ 0&0&1&0&1\\ 0&0&1&1&1\end{bmatrix}, \matr{T}\defeq \matr{H}_{[4]\setminus  1} =\begin{bmatrix} 1&1&0&0\\1&0&1&0\\ 0&1&0&1\\ 0&1&1&1\end{bmatrix} $$ and 
$$ \matr{T}_{[4]\setminus 1, [4]\setminus  1} =\begin{bmatrix}0&1&0\\ 1&0&1\\ 1&1&1\end{bmatrix},
\matr{T}_{[4]\setminus 1, [4]\setminus  2} =\begin{bmatrix}1&1&0\\ 0&0&1\\ 0&1&1\end{bmatrix}$$

We can compute the Bethe permanents of the matrices $\matr{T}$ and $ \matr{T}_{[4]\setminus 1, [4]\setminus  1} $ as follows.
\begin{align*}&\perm\big( \matr{T}^{\uparrow\matr{P}}\big) =\perm\begin{bmatrix} I&I&0&0\\I&0&I&0\\ 0&I&0&I\\ 0&I&P&Q\end{bmatrix}=\perm(I+P+Q)\\&\perm \big(\matr{T}_{[4]\setminus 1, [4]\setminus  1} ^{\uparrow\matr{P}_{[4]\setminus 1, [4]\setminus  1}}\big)=\perm\begin{bmatrix}0&I&0\\ I&0&I\\ I&P&Q\end{bmatrix}=\perm(I+Q), \\&\perm\big(\matr{T}_{[4]\setminus 1, [4]\setminus  2} ^{\uparrow\matr{P}_{[4]\setminus 1, [4]\setminus  1}}\big)=\perm\begin{bmatrix}I&I&0\\ 0&0&I\\ 0&P&Q\end{bmatrix}=\perm(P).\end{align*}

The inequality $\vomega_{\rm B,1} \leq \vomega_{\rm B,2} +\vomega_{\rm B,3}$ becomes  equivalent to  $${\hat t}_{3,M}^{1/M}\leq 1+(M+1)^{1/M},$$ which can be shown to be true by upper bounding $\vomega_{\rm B,1}$ by $(2^{M+1}-1)^{1/M}$ which, in the limit, is less than equal to $1+(M+1)^{1/M}$, as we can see in the proof of Corollary \ref{3-4-0pseudocdewods}, (part 3).  Note that the inequality $\vomega_{\rm B,M,1} \leq \vomega_{\rm B,M, 2} +\vomega_{\rm B,M, 3}$ 
also holds as we can see by plotting ${\hat t}_{3,M}^{1/M}$ and $1+(M+1)^{1/M}$ in Maple. Proving the inequality directly does not seem easy.
\end{example}

%

\subsection{Bethe Pseudocodewords for $\matr{H}$ of Size $2\times 3$}\label{2-3-0-section}
Although the following two cases of  $\matr{H}$ of size $2\times 3$ presented in this section, and $3\times 4$ presented in Section \ref{3-4-0-section},
are simple, they allow us to appreciate the difficulty of computing the degree-$M$ Bethe permanent vectors, even for small cases and provide some guidance in such an  attempt.


\begin{corollary} \label{2-3pseudocodewords} The possible Bethe pseudocodewords for a  $2\times 3$ matrix $\matr{H}$ (with no zero row or column) are equivalent to the following vectors (see Def. \ref{def:pseudocodewords}).  \begin{enumerate}
\item If  $\matr{H}=\left(\begin{matrix}1&1&1\\1&1&1\end{matrix}\right),$ then
$\vomega_{\mathrm{B},M}\propto\vomega_{\mathrm{B}}=(1,1,1)$. 
\item If $\matr{H}=\left(\begin{matrix}1&1&1\\0&1&1\end{matrix}\right),$
then $\vomega_{\mathrm{B},M}=\left(1, 1, (M+1)^{1/M}\right)$ and $\vomega_{\mathrm{B}}=(1,1,1).$ 
\item   If  $\matr{H}=\left(\begin{matrix}1&0&1\\1&1&0\end{matrix}\right),$ then
$\vomega_{\mathrm{B},M}=\vomega_{\mathrm{B}}=(1,1,1)$.
\item If  $\matr{H}=\left(\begin{matrix}1&1&1\\1&0&0\end{matrix}\right)$ or $\matr{H}=\left(\begin{matrix}0&1&1\\1&0&0\end{matrix}\right),$ then\\
$\vomega_{\mathrm{B},M}=\vomega_{\mathrm{B}}=(0,1,1)$.
\end{enumerate}
\end{corollary}
\begin{IEEEproof} We apply Theorem \ref{2-3case} to compute the vectors in each case.\\
1) $\vomega_{\mathrm{B},M}=\left( (M+1)^{1/M}, (M+1)^{1/M}, (M+1)^{1/M}\right)\propto(1,1,1).$ Taking the limit when $M\rightarrow \infty$  we obtain  $$\vomega_{\mathrm{B}}=\lim\limits_{M\rightarrow \infty} \vomega_{B, M}=(1,1,1). $$ 
2)  $\vomega_{\mathrm{B},M}=\left( t_{2,M}^{1/M},t_{2,M}^{1/M}, q_{2,M}^{1/M}\right)=(1, 1, (M+1)^{1/M}),$
where $t_{2,M}\defeq {\left<\perm\begin{bmatrix}1&1\\1&0\end{bmatrix}^{\uparrow\matr{P}}\right>_{\matr{P}\in \Psi_{2,M}}}\!\!\!\!=\perm\begin{bmatrix} I &I\\ I& 0\end{bmatrix}= q_1=1.$

Note that $(M+1)^{1/M}\leq 2\Longleftrightarrow M+1\leq 2^M$, for all $M\geq 1$, 
proving  that  $\vomega_{\mathrm{B},M}$ is a pseudocodeword as the general theorem stated. It results in the Bethe perm-pseudocodeword $(1,1,1)$. 

Cases 3) and 4) follow similarly. 
\end{IEEEproof}  
\begin{remark}
For all  $M\geq 1$,  the vector $(1,1, (M+1)^{1/M})$  is always a pseudocodeword for $\matr{H}$ equal to the all-one  $2\times 3$ matrix. 
Its AWGNC-pseudo-weight is equal to $\frac{\left(2+(M+1)^{1/M}\right)^2}{2+(M+1)^{2/M}}$ which is an increasing function that has a minimum equal to $8/3$ and this is attained for $M=1$, giving the pseudocodeword $\vomega_{B,1}=(1,1,2)$. The Bethe perm-pseudocodeword $\vomega_{\mathrm{B}}=(1,1,1)$ has AWGNC-pseudo-weight equal to $3$. The perm-pseudocodeword  $\vomega_{B,1}=(1,1,2)$ is more valuable than the Bethe perm-pseudocodeword $\vomega_{\mathrm{B}}=(1,1,1)$ because it gives a better bound on the minimum pseudo-weight. This is to be expected. The importance of the Bethe permanents is visible only for large matrices, where the algorithm to compute the permanent fails due to its high complexity. 
\end{remark}

\subsection{Bethe Pseudocodewords for $\matr{H}$ of Size $3\times 4$}\label{3-4-0-section}

We obtain the following corollary. 
\begin{corollary} \label{3-4-0pseudocdewods} The Bethe pseudocodewords for the  non-trivial $3\times 4$ matrices $\matr{H}$ (with no zero row or column)  are the following.  

1) If $\matr{H}=\left(\begin{matrix}1&1&1&1\\0&1&1&1\\0&1&1&1\end{matrix}\right),$ 
then\\
$\vomega_{\mathrm{B},M}=\left(q_{3,M}^{1/M},(M+1)^{1/M}, (M+1)^{1/M},(M+1)^{1/M}\right).$

2) If  $\matr{H}=\left(\begin{matrix}1&1&1&0\\0&1&1&1\\0&1&1&1\end{matrix}\right)$, then\\
$\vomega_{\mathrm{B},M}=\left(t_{3,M}^{1/M}, (M+1)^{1/M}, (M+1)^{1/M}, (M+1)^{1/M}\right).$

3) If  $\matr{H}=\left(\begin{matrix}1&1&1&0\\0&1&0&1\\0&1&1&1\end{matrix}\right)$, then\\
$\vomega_{\mathrm{B},M}=\left({\hat t}_{3,M}^{1/M}, 1, (M+1)^{1/M}, 1 \right).$

4) If  $\matr{H}=\left(\begin{matrix}1&1&1&0\\0&1&0&1\\0&0&1&1\end{matrix}\right)$, then\\
$\vomega_{\mathrm{B},M}=\left((M+1)^{1/M}, 1,1,1\right).$

\end{corollary}

\begin{IEEEproof} 
1) Shown for the general case, it is equivalent to  $q_{3,M}^{1/M}\leq 3(M+1)^{1/M}=3q_{2,M}^{1/M}$.

2) Observing that:\\\vspace{-3mm} \begin{center} ${{ \sum\limits_{r=0}^{M} {M \choose r} \sum\limits_{s=0}^{r} {r \choose s}(M-r+s)!(M-s)! \sum\limits_{t=0}^{M-r} {M-r \choose t}(M-t)! (r+t)!}\over{M!^3}}$\end{center}\vspace{-3mm}
{\small\begin{align*}
 \leq&{{\sum\limits_{r=0}^{M} {M \choose r} \sum\limits_{s=0}^{r} {r \choose s} (M-r)!M!r!}\over{M!^2}}\\=&
 {{\sum\limits_{r=0}^{M} {M \choose r} (M-r)!M!M!r!\sum\limits_{s=0}^{r} {r \choose s}\sum\limits_{t=0}^{M-r} {M-r \choose t}}\over{M!^3}}\\=&
 \sum\limits_{r=0}^{M}  \sum\limits_{s=0}^{r} {r \choose s}\sum\limits_{t=0}^{M-r} {M-r \choose t}
  \sum\limits_{r=0}^{M} 2^r\cdot 2^{M-r}\\= & 2^M\sum\limits_{r=0}^{M} 1=2^M(M+1),  \end{align*}}\noindent
 yields that $\vomega_{\mathrm{B},M}$ is a pseudocodeword.  Taking the limit, we get that the Bethe permanent vector $\vomega_{\mathrm{B}}$ is also a pseudocodeword. 

3) Plotting ${\hat t}_{3,M}^{1/M}$ and $1+(M+1)^{1/M}$ we see that ${\hat t}_{3,M}^{1/M}\leq 1+(M+1)^{1/M}$. \\
Note that we can prove that the Bethe permanent is a pseudocodeword  directly, without proving that the degree-$M$ Bethe permanent is s pseudocodeword by observing that 
\begin{align*}{\hat t}_{3,M}=&{{\sum\limits_{r=0}^{M} {M \choose r} \sum\limits_{s=0}^{r} {r \choose s}(M-r+s)!(M-s)! r!}\over{M!^2}}\\
 \leq&{{\sum\limits_{r=0}^{M} {M \choose r} \sum\limits_{s=0}^{r} {r \choose s} (M-r)!M!r!}\over{M!^2}}\\=&
 {{\sum\limits_{r=0}^{M}  \sum\limits_{s=0}^{r} {r \choose s}}}=
 \sum\limits_{r=0}^{M}  2^r=  2^{M+1}-1. \end{align*}
Taking the limit, we obtain   $\lim_{M\rightarrow \infty}{\hat t}_{3,M}\leq (2^{M+1}-1)^{1/M}=2\leq 1+1=1+\lim_{M\rightarrow \infty}({M+1})^{1/M}$, so we can deduce computationally as well that the Bethe permanent vector is a pseudocodeword. 

4) The inequality $(M+1)^{1/M}\leq 2$ is true for any $M\geq 1$.     
\end{IEEEproof}

\begin{example}
  The vectors $(4,2,2,2)\propto (2,1,1,1)$ for $M=1$ and,  e.g.,  $(\sqrt{10}, \sqrt{3},\sqrt{3},\sqrt{3})$
for $M=2$ for the matrix $\matr{H}$ in  part 2)  of Cor. \ref{3-4-0pseudocdewods} are two examples of  pseudocodewords obtained with the techniques described above. 
 \end{example}

Note that sometimes we can obtain valid pseudocodewords by taking block-permanent vectors based on a set of size $m+1$ (without averaging over all possible lifts) as the Example \ref{example-motivation} showed. Indeed, the vector $ (48^{1/M},  2^{1/M},2^{1/M},2^{1/M})=\sqrt[3]{2}(2\sqrt[3]{3},1,1,1)$ obtained there was obtained in this way. It is a pseudocodeword equivalent to the pseudocodeword $(2\sqrt[3]{3},1,1,1)$ with AWGNC-pseudo-weight $3.58$. Note that this is not always the case, an example can be easily found. In general, averages over all liftings {\it need} to be taken in order to obtain pseudocodewords.

\section{Concluding Remarks} \label{sec:conclusions}

We summarize our results in the following paragraphs and offer some suggestions  for future directions in attempting the open problems.

  We showed that it is enough to prove the conjecture for matrices with the first column of weight one and to  only show one inequality of the fundamental cone, $$\vomega_{{\rm B},M,1}\leq \sum_{i\in \supp(\vect{R}_1)\setminus 1} \vomega_{{\rm B},M,i},$$
 instead of $\sum_{i}^m \supp(\vect{R}_i)$ inequalities.   
 
 We also showed that the conjecture is equivalent to a statement about  the Bethe permanent of a square matrix 
 $\matr{T}=(t_{ij})_{1\leq i,j\leq m}$ being less than or equal to the Bethe permanent-expansion on the first row:
 $$\perm_{\mathrm{B}}\big( \matr{T}\big) \leq  \sum_{l\in[m]}
           t_{1l}
           \cdot
           \perm_{\mathrm{B}}\big( \matr{T}_{[m]\setminus 1, [m]\setminus  l} \big).$$  
 
  We proved a  stronger form of the conjecture, which states that the degree-$M$ Bethe permanent vectors are pseudocodewords, for two families of matrices. By taking the limit,  the conjecture's claim  that the Bethe permanent vectors are pseudocodewords follows for these familes of matrices.  

 We also discussed a natural next  case to consider and showed how we can approach it. This case cannot be solved in the way we solved the two cases. When proving the inequality $$\vomega_{{\rm B},M,1}\leq \sum_{i\in \supp(\vect{R}_1)\setminus 1} \vomega_{{\rm B},M,i}, $$  if the terms on the right are not all equal (as they are in the two solved cases) then our method fails and  a different idea is needed.  A possible direction in solving this case  is to show instead that $$\vomega_{{\rm B},M,1}\leq \left(|\supp(\vect{R}_1)|- 1\right)|\cdot \min_{i\in \supp(\vect{R}_1)\setminus 1} \vomega_{{\rm B},M,i} $$
 which is equivalent to showing  the simpler expression\footnote{The $M$ power annihilates the $M$th root in the definition of the degree $M$ Bethe permanents. The expression becomes an inequality between averages of permanents of matrices.} 
 $$\left(\vomega_{{\rm B},M,1}\right)^M\leq\left( |\supp(\vect{R}_1)|-1\right)^M\!\!\cdot\! \left(\min_{i\in \supp(\vect{R}_1)\setminus 1} \vomega_{{\rm B},M,i} \right)^M$$ and then use arguments similar to our solved cases. Unfortunately, we found an example for which this is not true. It is true, however, at the limit, so we conjecture that this is always the case.   
 
 A different direction that we attempted successfully in several cases is  to compute the  degree-$M$ Bethe   permanents  exactly,  using  combinatorial arguments. This might be possible for matrices with a lot of zeros (like the ones we usually encounter) by reducing  the computation of the permanent of a sparse block $mM\times mM$ matrix to that of the permanent of a smaller matrix and to use the closed forms of the Bethe permanents and degree-$M$ Bethe permanents  of small matrices calculated in this paper.  We illustrated this in an example. 

\section*{Acknowledgements} We would like to thank Pascal O. Vontobel for suggesting this problem and for our many discussions on this topic and Martin Haenggi for his valuable help with  the related programming and the numerical investigation. 
  
\appendix\label{matrixgeneral}
\bigformulatop{\value{equation}}
{\begin{align}& 
T_{\beta}=
\begin{bmatrix}
 \left.\begin{matrix}\begin{matrix} \cdots &\begin{matrix} 1 \\0\\\vdots\\0\end{matrix}&\cdots \end{matrix}  \\A^{(1)}_{\beta_1} \end{matrix} \right| & 
 \left.\begin{matrix}\begin{matrix} \cdots &\begin{matrix} 1 \\0\\\vdots\\0\end{matrix}&\cdots \end{matrix}  \\A^{(2)}_{\beta_2} \end{matrix} \right| &
 \left.\begin{matrix} \begin{matrix} \cdots &\begin{matrix} 1 \\ 0\\ \vdots\\0 \end{matrix}& \cdots &\begin{matrix} 0 \\1\\ \vdots\\0\end{matrix}&\cdots \end{matrix} \\ A^{(3)}_{\beta_3}\end{matrix}  \right|&\cdots &
\left|\begin{matrix} \begin{matrix} \cdots &\begin{matrix} 1 \\ 0\\ \vdots\\0 \end{matrix}& \cdots &\begin{matrix} 0 \\1\\ \vdots\\0\end{matrix}&\cdots \end{matrix} \\ A^{(m)}_{\beta_m}\end{matrix} \right.
\end{bmatrix}\label{formulatop1}\\
~\nonumber\\
&T'_{\beta}=\begin{bmatrix}
 \left.\begin{matrix}\begin{matrix} \cdots &\begin{matrix} 1 \\0\\\vdots\\0\end{matrix}&\cdots \end{matrix}  \\A^{(1)}_{\beta_1} \end{matrix} \right| & 
 \left.\begin{matrix}\begin{matrix} \cdots &\begin{matrix} 0 \\1\\\vdots\\0\end{matrix}&\cdots \end{matrix}  \\A^{(2)}_{\beta_2} \end{matrix} \right| &
 \left.\begin{matrix} \begin{matrix} \cdots &\begin{matrix} 1 \\ 0\\ \vdots\\0 \end{matrix}& \cdots &\begin{matrix} 0 \\1\\ \vdots\\0\end{matrix}&\cdots \end{matrix} \\ A^{(3)}_{\beta_3}\end{matrix}  \right|&\cdots &
\left|\begin{matrix} \begin{matrix} \cdots &\begin{matrix} 1 \\ 0\\ \vdots\\0 \end{matrix}& \cdots &\begin{matrix} 0 \\1\\ \vdots\\0\end{matrix}&\cdots \end{matrix} \\ A^{(m)}_{\beta_m}\end{matrix} \right.
\end{bmatrix}\label{formulatop2}
\\~\nonumber\\
&\perm (T_{\beta})-\perm (T'_{\beta})\nonumber =\\&
\perm \begin{bmatrix}
 \left.\begin{matrix}\begin{matrix} \cdots &\begin{matrix} 0 \\0\\\vdots\\0\end{matrix}&\cdots \end{matrix}  \\A^{(1)}_{\beta_1} \end{matrix} \right| & 
 \left.\begin{matrix}\begin{matrix} \cdots &\begin{matrix} 1 \\0\\\vdots\\0\end{matrix}&\cdots \end{matrix}  \\A^{(2)}_{\beta_2} \end{matrix} \right| &
 \left.\begin{matrix} \begin{matrix} \cdots &\begin{matrix} 0 \\ 0\\ \vdots\\0 \end{matrix}& \cdots &\begin{matrix} 0 \\1\\ \vdots\\0\end{matrix}&\cdots \end{matrix} \\ A^{(3)}_{\beta_3}\end{matrix}  \right|&\cdots &
\left|\begin{matrix} \begin{matrix} \cdots &\begin{matrix} 0 \\ 0\\ \vdots\\0 \end{matrix}& \cdots &\begin{matrix} 0 \\1\\ \vdots\\0\end{matrix}&\cdots \end{matrix} \\ A^{(m)}_{\beta_m}\end{matrix} \right.
\end{bmatrix}\nonumber -  \\&
\perm\begin{bmatrix}
 \left.\begin{matrix}\begin{matrix} \cdots &\begin{matrix} 1 \\0\\\vdots\\0\end{matrix}&\cdots \end{matrix}  \\A^{(1)}_{\beta_1} \end{matrix} \right| & 
 \left.\begin{matrix}\begin{matrix} \cdots &\begin{matrix} 0 \\1\\\vdots\\0\end{matrix}&\cdots \end{matrix}  \\A^{(2)}_{\beta_2} \end{matrix} \right| &
 \left.\begin{matrix} \begin{matrix} \cdots &\begin{matrix} 1 \\ 0\\ \vdots\\0 \end{matrix}& \cdots &\begin{matrix} 0 \\0\\ \vdots\\0\end{matrix}&\cdots \end{matrix} \\ A^{(3)}_{\beta_3}\end{matrix}  \right|&\cdots &
\left|\begin{matrix} \begin{matrix} \cdots &\begin{matrix} 1 \\ 0\\ \vdots\\0 \end{matrix}& \cdots &\begin{matrix} 0 \\0\\ \vdots\\0\end{matrix}&\cdots \end{matrix} \\ A^{(m)}_{\beta_m}\end{matrix} \right.
\end{bmatrix}\label{formulatop3}
\\~\nonumber\\
&\perm (T_{\beta})-\perm (T'_{\beta})\nonumber\\=&
\perm \begin{bmatrix}
\left.\begin{matrix}\begin{matrix} \cdots &\begin{matrix} 0 \\0\\\vdots\\0\end{matrix}&\cdots \end{matrix}  \\A^{(1)}_{\beta_1} \end{matrix} \right| & 
 \left.\begin{matrix}\begin{matrix} \cdots &\begin{matrix} 1 \\0\\\vdots\\0\end{matrix}&\cdots \end{matrix}  \\A^{(2)}_{\beta_2} \end{matrix} \right| &
 \left.\begin{matrix} \begin{matrix} \cdots &\begin{matrix} 0 \\ 0\\ \vdots\\0 \end{matrix}& \cdots &\begin{matrix} 0 \\1\\ \vdots\\0\end{matrix}&\cdots \end{matrix} \\ A^{(3)}_{\beta_3}\end{matrix}  \right|&\cdots &
\left|\begin{matrix} \begin{matrix} \cdots &\begin{matrix} 0 \\ 0\\ \vdots\\0 \end{matrix}& \cdots &\begin{matrix} 0 \\1\\ \vdots\\0\end{matrix}&\cdots \end{matrix} \\ A^{(m)}_{\beta_m}\end{matrix} \right.
\end{bmatrix}\nonumber -  \\&
\perm\begin{bmatrix}
 \left.\begin{matrix}\begin{matrix} \cdots &\begin{matrix} 0 \\1\\\vdots\\0\end{matrix}&\cdots \end{matrix}  \\A^{(1)}_{\beta_1} \end{matrix} \right| & 
 \left.\begin{matrix}\begin{matrix} \cdots &\begin{matrix} 1 \\0\\\vdots\\0\end{matrix}&\cdots \end{matrix}  \\A^{(2)}_{\beta_2} \end{matrix} \right| &
 \left.\begin{matrix} \begin{matrix} \cdots &\begin{matrix} 0 \\ 1\\ \vdots\\0 \end{matrix}& \cdots &\begin{matrix} 0 \\0\\ \vdots\\0\end{matrix}&\cdots \end{matrix} \\ A^{(3)}_{\beta_3}\end{matrix}  \right|&\cdots &
\left|\begin{matrix} \begin{matrix} \cdots &\begin{matrix} 0 \\ 1\\ \vdots\\0 \end{matrix}& \cdots &\begin{matrix} 0 \\0\\ \vdots\\0\end{matrix}&\cdots \end{matrix} \\ A^{(m)}_{\beta_m}\end{matrix} \right.
\end{bmatrix}\leq\label{formulatop4}
\\&\perm \begin{bmatrix}
\left.\begin{matrix}\begin{matrix} \cdots &\begin{matrix} 0 \\1\\\vdots\\0\end{matrix}&\cdots \end{matrix}  \\A^{(1)}_{\beta_1} \end{matrix} \right| & 
 \left.\begin{matrix}\begin{matrix} \cdots &\begin{matrix} 1 \\0\\\vdots\\0\end{matrix}&\cdots \end{matrix}  \\A^{(2)}_{\beta_2} \end{matrix} \right| &
 \left.\begin{matrix} \begin{matrix} \cdots &\begin{matrix} 0 \\ 0\\ \vdots\\0 \end{matrix}& \cdots &\begin{matrix} 0 \\1\\ \vdots\\0\end{matrix}&\cdots \end{matrix} \\ A^{(3)}_{\beta_3}\end{matrix}  \right|&\cdots &
\left|\begin{matrix} \begin{matrix} \cdots &\begin{matrix} 0 \\ 0\\ \vdots\\0 \end{matrix}& \cdots &\begin{matrix} 0 \\1\\ \vdots\\0\end{matrix}&\cdots \end{matrix} \\ A^{(m)}_{\beta_m}\end{matrix} \right.
\end{bmatrix} \nonumber-  \\&\perm \begin{bmatrix}
\left.\begin{matrix}\begin{matrix} \cdots &\begin{matrix} 0 \\1\\\vdots\\0\end{matrix}&\cdots \end{matrix}  \\A^{(1)}_{\beta_1} \end{matrix} \right| & 
 \left.\begin{matrix}\begin{matrix} \cdots &\begin{matrix} 1 \\0\\\vdots\\0\end{matrix}&\cdots \end{matrix}  \\A^{(2)}_{\beta_2} \end{matrix} \right| &
 \left.\begin{matrix} \begin{matrix} \cdots &\begin{matrix} 0 \\ 1\\ \vdots\\0 \end{matrix}& \cdots &\begin{matrix} 0 \\0\\ \vdots\\0\end{matrix}&\cdots \end{matrix} \\ A^{(3)}_{\beta_3}\end{matrix}  \right|&\cdots &
\left|\begin{matrix} \begin{matrix} \cdots &\begin{matrix} 0 \\ 1\\ \vdots\\0 \end{matrix}& \cdots &\begin{matrix} 0 \\0\\ \vdots\\0\end{matrix}&\cdots \end{matrix} \\ A^{(m)}_{\beta_m}\end{matrix} \right.
\end{bmatrix}\label{formulatop5}
\end{align}
}
\addtocounter{equation}{5}
\subsection{Proof  of Lemma \ref{recursive1}} \label{proof:lemma}
 Let $\beta\defeq \{\beta_1,\ldots, \beta_m\}$, and let 
\begin{align*}T_{\beta}\defeq &\begin{bmatrix}  Q_{11,\beta_1}& Q_{12,\beta_2}&\cdots &Q_{1m, \beta_m}\\ \vdots&\vdots&&\vdots\\Q_{m-1,1, \beta_1}&Q_{m-1,2,\beta_2}&\cdots &Q_{m-1,m, \beta_m}\end{bmatrix}\end{align*}
like in the statement of Lemma \ref{recursive1}. Since $\{[M]\setminus\beta_i\}_{i=1}^{i=M}$ is a partition of $[M]$, then $\sum\limits_i \lvert\beta_i \rvert = (m-1)\cdot M$.
Therefore, $T_{\beta}$ is of size $(m-1)M \times (m-1)M$. Each of the columns of $\begin{bmatrix} Q_{11,\beta_1}& Q_{12,\beta_2}&\cdots &Q_{1m, \beta_m}\end{bmatrix}$ has Hamming weight 1 and each of its rows has Hamming weight at most $m$. 
 Let $n_0, n_1,\ldots, n_{m}$ be the number of  rows of weight $0, 1,  \ldots, m$, respectively. 
 The total number of rows is then $n_0+n_1+n_2+\cdots +n_{m}= M$ and the total number of entries equal to 1 in the matrix is $n_1+2n_2+\cdots +mn_{m}=(m-1)M$. We obtain:\\
$(m-1)\cdot (n_0+n_1+ n_2+\cdots +n_m)=n_1+2n_2+\cdots +mn_{m} \Leftrightarrow \\
(m-1)n_0+(m-2)n_1+ (m-3)n_2+\cdots +n_{m-2}  =n_{m}.
$
 If $n_m>0$ we obtain that with each row of weight $m$ there is at least one row of weight $m-2$ or lower.  If $n_m=0$ we obtain that 
$n_0=n_1=\cdots =n_{m-2} =0$ and hence all rows and columns have constant Hamming weight $m-1$.  We will show that the first case can be reduced to the second case by a modification of the matrix that leaves the permanent unchanged or increases it.

We have two cases that we prove independently, the case of   $n_m>0$ and the case of $n_m=0$. The proof of the first case is followed by Example \ref{example1} and the proof of the second case contains Example \ref{example2} and is followed by Example \ref{example3} in order to better illustrate the mathematical techniques used in this proof.

\fbox{\bf Case $n_m>0$.} \begin{IEEEproof} We can assume that $n_0=0$, otherwise the permanent is 0 and does not contribute to the sum value of all permanents. 
There are $m$ permutation matrices in the matrix $\begin{bmatrix} Q_{11}& Q_{12}&\cdots &Q_{1m}\end{bmatrix}$ from which columns indexed by $\beta_1, \beta_2, \ldots,$ and $\beta_m$, respectively, are chosen to give the $M\times M(m-1)$ matrix $\begin{bmatrix} Q_{11,\beta_1}& Q_{12,\beta_2}&\cdots &Q_{1m, \beta_m}\end{bmatrix}$.  Therefore, the inequality $n_m>0$ means that there exists a column $\begin{bmatrix} 0&\cdots&0&1&0&\cdots&0\end{bmatrix}^\tr$ that gets picked from each of the matrices $ Q_{11}, Q_{12},\cdots, Q_{1m}$. This gives  $\beta_i>0$, for all $i\in [m]$.  We can permute the first $M$ rows of $T_\beta$ (process that does not alter its permanent) and assume that this column is $\begin{bmatrix} 1&0&\cdots&0\end{bmatrix}^\tr$. We also obtain that there is another column which gets picked from at most $m-2$ of the matrices $Q_{11}, Q_{12},\ldots, Q_{1m}$.  We permute the first $M$ rows of $T_\beta$ by leaving the first row fixed, and assume that this column is $\begin{bmatrix} 0&1&0&\cdots&0\end{bmatrix}^\tr$. We can also assume that this column does not appear in $Q_{11,\beta_1}$ and  $Q_{12,\beta_2}$. 
Therefore, without loss of generality, we can assume that $T_{\beta}$ is of the form displayed in \eqref{formulatop1}, where, (up to column permutations within each block of $M$ columns) 
\begin{align*} &Q_{11,\beta_1}= \begin{bmatrix} \cdots &\begin{matrix} 1 \\0\\\vdots\\0\end{matrix}&\cdots \end{bmatrix},\quad
Q_{12,\beta_2}= \begin{bmatrix} \cdots &\begin{matrix} 1 \\0\\\vdots\\0\end{matrix}&\cdots \end{bmatrix},\end{align*}
\begin{align*}& Q_{1j, \beta_j}= \begin{bmatrix} \cdots &\begin{matrix} 1 \\ 0\\ \vdots\\0 \end{matrix}& \cdots &\begin{matrix} 0 \\1\\ \vdots\\0\end{matrix}&\cdots \end{bmatrix}, \quad j\geq 3,\end{align*}
and 
\begin{align*} &\begin{bmatrix} A^{(1)}_{\beta_1} & A^{(2)}_{\beta_2}&\cdots &A^{(m)}_{\beta_m}\end{bmatrix}\\ &\defeq\begin{bmatrix}  Q_{21,\beta_1}& Q_{22,\beta_2}&\cdots &Q_{2m, \beta_m}\\ \vdots&\vdots&&\vdots\\ Q_{m-1,1,\beta_1}&Q_{m-1,2,\beta_2}&\cdots &Q_{m-1,m, \beta_m}\end{bmatrix}.\end{align*}

Let $T'_\beta$ be the matrix displayed in \eqref{formulatop2}. The two matrices $ T_{\beta}$ and $ T'_{\beta}$ differ in two positions only, in the first two  rows and the second block of $M$ columns. An entry equal to 1 in the first row gets ``moved" to the position below in the second row and the same column. Therefore, the permanents of $T_{\beta}$ and $ T'_{\beta}$ differ only in the elementary products (products containing exactly one entry from each row and each column) that contain the ``moving 1 entry"; their difference is given in \eqref{formulatop3}.  The first matrix in \eqref{formulatop3} has its first row of weight $1$ (changed from the first row of $T_\beta$ of weight $m$)  and its second row of weight up to $m-2$ (equal to the second row of $T_\beta$).  The second matrix in \eqref{formulatop3} has its first row of weight $m-1$ (changed from the first row of  $T'_\beta$ of weight $m$)  and the second row of weight $1$  (changed from the second row of $T'_\beta$ of weight less than or equal to  $m-1$). Permuting the first two rows of the second matrix in \eqref{formulatop3} gives formula \eqref{formulatop4} and changing a zero entry into a 1 entry on the second row of the first matrix gives a larger difference of permanents expressed in  \eqref{formulatop5}.

 \bigformulatop{\value{equation}}
{\begin{align} T_\beta&\defeq \begin{bmatrix} Q_{11,\beta_1}& Q_{12,\beta_2}&Q_{13,\beta_3}\\Q_{21,\beta_1}& Q_{22,\beta_2}&Q_{23,\beta_3}\end{bmatrix} \defeq \begin{bmatrix} \begin{matrix} 1&0\\0&0\\0&1\end{matrix}&~&\begin{matrix} 0&1\\1&0\\0&0\end{matrix}&~&\begin{matrix} 0&0 \\0&1\\1&0\end{matrix}\\ ~ \\ \begin{matrix} 0&0\\0&1\\1&0\end{matrix}&~&\begin{matrix} 1&0\\0&0\\0&1\end{matrix}&~&\begin{matrix} 1&0\\0&1\\0&0\end{matrix}\end{bmatrix},\label{matrixexample1} \\
 \setS_{12}&=\begin{bmatrix} 0\\0\\1\end{bmatrix}, \quad \setS_{13}=\begin{bmatrix} 1\\0\\0\end{bmatrix}, \quad \setS_{22}=\begin{bmatrix} 0\\1\\0\end{bmatrix},\quad  \setS_{23}=\begin{bmatrix} 0\\0\\1\end{bmatrix} \label{matrixexample2}\\
 P_{11}&\defeq \begin{bmatrix}Q_{12,\beta_2}&\setS_{12} \end{bmatrix}=\begin{bmatrix} 0&1&0\\1&0&0\\0&0&1\end{bmatrix},\quad
 P_{12}\defeq \begin{bmatrix}Q_{13,\beta_3}&\setS_{13} \end{bmatrix}=\begin{bmatrix}0&0&1\\0&1&0\\1&0&0\end{bmatrix},\nonumber\\
 P_{21}&\defeq \begin{bmatrix}Q_{22,\beta_2}&\setS_{22} \end{bmatrix}=\begin{bmatrix} 1&0&0\\0&0&1\\0&1&0\end{bmatrix},\quad
P_{22}\defeq \begin{bmatrix}Q_{23,\beta_3}&\setS_{23} \end{bmatrix}=\begin{bmatrix}1&0&0\\0&1&0\\0&0&1\end{bmatrix}.\label{permutationsexample2}\end{align}}
\addtocounter{equation}{3}
 last two matrices in  \eqref{formulatop5} differ now only in the second row, namely, in  the positions of the 1 entries on the second row in the last $m-2$ blocks.  They have both the first row of weight 1, the second row of weight $m-1$, and all the other rows equal to the corresponding rows of $T_\beta$. We now observe that allowing the matrices $Q_{ij}$ with $2\leq i \leq m-1$ and $j\in[m]$,  to vary among all possible permutation matrices, while keeping the first $M$ rows fixed, gives us two equal sets of matrices.  This can be seen by interchanging two columns in each $Q_{ij}$ of a matrix in the first set to obtain a permutation  matrix $Q'_{ij}$,  for all $2\leq i \leq m-1$ and $3\leq j \leq m$. Setting also $Q'_{ij}=Q_{ij}$,  $2\leq i\leq m-1, j\in \{1,2\}$, then these choices of $Q'_{ij}$ with $i\geq 2$ and $j\geq 1$ give a matrix in the second set, and vice-versa. The two sets being equal means that the sum of all differences of their permanents over all matrices $Q_{ij}$ with $2\leq i \leq m-1$ and $j\in[m]$ is $0$. Since  
 \begin{align*}
&\sum\limits_{\tiny\begin{matrix}Q_{ij}\in \mathcal{P}_M\\ i\in [m-1]\setminus \{1\}\\j\in [m]\end{matrix}}\perm(T_{\beta})-\perm(T'_{\beta})  \leq 0\end{align*} 
 we obtain   
\begin{align*}
&\sum\limits_{\tiny\begin{matrix}Q_{ij}\in \mathcal{P}_M\\ i\in [m-1]\\j\in [m]\end{matrix}}\perm(T_{\beta})=\sum\limits_{\tiny\begin{matrix}Q_{1j}\in \mathcal{P}_M\\  j\in [m]\end{matrix}}\sum\limits_{\tiny\begin{matrix}Q_{ij}\in \mathcal{P}_M\\  i\in [m-1]\setminus \{1\}\end{matrix}}\perm(T_{\beta}) \\\leq
&\sum\limits_{\tiny\begin{matrix}Q_{1j}\in \mathcal{P}_M\\  j\in [m]\end{matrix}} \sum\limits_{\tiny\begin{matrix}Q_{ij}\in \mathcal{P}_M\\i\in [m-1]\setminus \{1\}\end{matrix}}\perm(T'_{\beta}) 
= \sum\limits_{\tiny\begin{matrix}Q_{ij}\in \mathcal{P}_M\\  i\in [m-1]\\j\in [m]\end{matrix}}\perm(T'_{\beta})
\end{align*}

 Therefore, by ``moving" a 1 entry from a row of weight $m$ (among the first $M$ rows)  to a row of weight at most $m-2$ (see \eqref {formulatop1} and \eqref{formulatop2}) gives a new matrix $T'_\beta$ {\em of the same form as the original matrix}, but with its first $M$ rows with only $n_m-1$ rows of weight $m$. Summing over all possible permutation matrices, this change gives a larger value of the permanent sum, therefore we can work with this new matrix. 
 
 We proceed similarly with each block of $M$ rows that contains a row of weight $m$, to obtain a matrix with each row of weight $m-1$.  Thus, we substitute each matrix in the first sum in \eqref{recursive-ineq1} having $n_m>0$ with a matrix that has  each row of weight $m-1$. \end{IEEEproof} 
 \begin{example} \label{example1} Let $M=3$,  $m=3$, 
 $$\matr{Q}\defeq\begin{bmatrix} \begin{matrix} 1&0&0\\0&1&0\\0&0&1\end{matrix}&~&\begin{matrix} 0&1&0\\1&0&0\\0&0&1\end{matrix}&~&\begin{matrix} 0&0&1\\0&1&0\\1&0&0\end{matrix}\\ ~ \\ \begin{matrix} 0&1&0\\0&0&1\\1&0&0\end{matrix}&~&\begin{matrix} 1&0&0\\0&0&1\\0&1&0\end{matrix}&~&\begin{matrix} 1&0&0\\0&1&0\\0&0&1\end{matrix}\end{bmatrix}$$
 and $\beta_1\defeq \{1,2\}, \beta_2\defeq \{1,2\}, \beta_3\defeq \{1,2\}$.
 Then $$T_\beta=\begin{bmatrix} \begin{matrix} 1&0\\0&1\\0&0\end{matrix}&~&\begin{matrix} 0&1\\1&0\\0&0\end{matrix}&~&\begin{matrix} 0&0
 \\0&1\\1&0\end{matrix}\\ ~ \\ \begin{matrix} 0&1\\0&0\\1&0\end{matrix}&~&\begin{matrix} 1&0\\0&0\\0&1\end{matrix}&~&\begin{matrix} 1&0\\0&1\\0&0\end{matrix}\end{bmatrix}.$$
 The second and the third rows of the first block of 3 rows of $T_\beta$ have weight $3=m$ and weight $1=m-2$, respectively. 
 Similarly, the first  and the second rows of the second  block of 3 rows of $T_\beta$ have weight $3=m$ and weight $1=m-2$, respectively.
 We can modify the matrix in two steps, by constructing   
 $$T'_\beta=\begin{bmatrix} \begin{matrix} 1&0\\0&0\\0&1\end{matrix}&~&\begin{matrix} 0&1\\1&0\\0&0\end{matrix}&~&\begin{matrix} 0&0
 \\0&1\\1&0\end{matrix}\\ ~ \\ \begin{matrix} 0&1\\0&0\\1&0\end{matrix}&~&\begin{matrix} 1&0\\0&0\\0&1\end{matrix}&~&\begin{matrix} 1&0\\0&1\\0&0\end{matrix}\end{bmatrix},$$
followed by constructing
  $$T''_\beta=\begin{bmatrix} \begin{matrix} 1&0\\0&0\\0&1\end{matrix}&~&\begin{matrix} 0&1\\1&0\\0&0\end{matrix}&~&\begin{matrix} 0&0
 \\0&1\\1&0\end{matrix}\\ ~ \\ \begin{matrix} 0&0\\0&1\\1&0\end{matrix}&~&\begin{matrix} 1&0\\0&0\\0&1\end{matrix}&~&\begin{matrix} 1&0\\0&1\\0&0\end{matrix}\end{bmatrix},$$
  and showing that the change from $T_\beta$ into $T'_\beta$ and then into $T''_\beta$ only increases the average permanent over all permutation matrices $\matr{Q}\in \Psi_{2,3,3}$. $T'_\beta$ has the first block of $M=3$ rows of weight $2=m-1$, while  $T''_\beta$ has constant  row weight $2=m-1$. 
 \end{example}
 
\fbox{\bf Case $n_m=0$.} \begin{IEEEproof}
In this case, we can split each of the matrices $Q_{i1,\beta_1}$, $i\in [m-1]$,  into $m-1$ sets $\setS_{ij}$ of columns such that the matrices 
$\begin{bmatrix}Q_{ij,\beta_j}&\setS_{ij} \end{bmatrix}\defeq P_{i,j-1}$ are all permutation matrices of size $M$, for all $i\in [m-1]$, and $2\leq j\leq m$. In other words,  each matrix $Q_{ij,\beta_j}$ with $ j\geq 2$  needs $M-|\beta_j|$ columns from the set  $\setS_{ij}$ to get ``completed"  to a permutation matrix $P_{i,j-1}$, and these columns may all be found in the matrix $Q_{i1,\beta_1}$, $i\in [m-1]$. Note that no one column in $Q_{i1,\beta_1}$, $i\in [m-1]$ can be in two different sets $\setS_{ij}$ and $\setS_{ik}$, $j\neq k$, or, equivalently, no one column is missing from two different matrices $Q_{ij,\beta_j}$, $Q_{ik,\beta_k}$, $j\neq k$, because if so, then there would exist a row with a lower weight than $m-1$, thus contradicting the assumption we are under in this case. 

Therefore, we can apply a permutation on the first columns in $\beta_1$ of $T_\beta$ to arrange them so that the first $M-|\beta_2|$ columns of these form the set $\setS_{12}$ that completes the matrix $Q_{12,\beta_2}$ to a permutation matrix $P_{11}$, the next $M-|\beta_3|$ columns of these form the set $\setS_{13}$ that completes the matrix $Q_{13,\beta_3}$ to a permutation matrix $P_{12}$, and so on. Since the permanent is not changed when permutations of columns or rows are applied, we can assume, without loss of generality,  that $T_\beta$ has this ``order"  on the first $\beta_1$ columns. 
We denote by $\sigma$ the permutation of columns of $T_\beta$ required to change its first $M$ rows as following (by extension, for simplicity,  $\sigma$ will also denote the induced permutation of columns on any submatrix of $T_\beta$ obtained by erasing some of its rows): 
 $$\sigma\begin{bmatrix}  Q_{11,\beta_1}&\cdots &Q_{1, m-1, \beta_m}\end{bmatrix}=\begin{bmatrix}  P_{11}&\cdots &P_{1, m-1}\end{bmatrix}.$$ 
Let $$\sigma(T_\beta) = \begin{bmatrix}  P_{11}&\cdots &P_{1, m-1}\\ R_{21}&\cdots &R_{2, m-1}\\
\vdots&\cdots&\vdots\\R_{m-1,1}&\cdots &R_{m-1, m-1} \end{bmatrix},$$ 
where each $R_{ij}= \begin{bmatrix}Q_{ij,\beta_j}&\setS_{1j}\end{bmatrix}$, $j\geq 2$, is a matrix of size $M\times M$.
If $R_{ij}$ are all permutation matrices, or equivalently, if $\setS_{ij}=\setS_{1j}$, then  
\begin{align}\label{smaller}\sigma(T_\beta) = \begin{bmatrix}  P_{11}&\cdots &P_{1, m-1}\\ P_{21}&\cdots &P_{2, m-1}\\
\vdots&\cdots&\vdots\\P_{m-1,1}&\cdots &P_{m-1, m-1} \end{bmatrix},\end{align}
\bigformulatop{\value{equation}}
{\begin{align}
&\sum\limits_{\tiny\begin{matrix}Q_{1j}, Q_{2j}\in \mathcal{P}_M\\j\in[m]\end{matrix}}\perm(T_\beta)-\perm(T'_\beta)\nonumber \\=&\sum\limits_{\tiny\begin{matrix}Q_{1j}, Q_{2j}\in \mathcal{P}_M\\j\in[m]\end{matrix}}
\perm\begin{bmatrix}   
Q_{11,\beta_1} &A_1\\
Q_{21,\beta_1}& A_2\\
 \vdots&\vdots\\ 
Q_{m-1,1,\beta_1} &A_{m-1}
\end{bmatrix}-
  \sum\limits_{\tiny\begin{matrix}Q_{1j}, Q_{2j}\in \mathcal{P}_M\\j\in[m]\end{matrix}}
 \perm\begin{bmatrix}  
Q_{11,\beta_1} & A_1\\
Q'_{21,\beta_1}&A_2\\
  \vdots&\vdots\\
 Q_{m-1,1,\beta_1}   &A_{m-1}
 \end{bmatrix}\nonumber\\\nonumber ~\\=
 &\sum\limits_{\tiny\begin{matrix}Q_{1j}, Q_{2j}\in \mathcal{P}_M\\j\in[m]\end{matrix}}
  \perm\begin{bmatrix}  
   Q_{11,\beta_1}  & A_1\\ ~\\
\left.\begin{matrix}\begin{matrix}1 &0\\0&1\\\vdots &\vdots \\0&0\end{matrix}&\cdots&\end{matrix}\right|&A_2\\
\vdots&  \vdots\\
Q_{m-1,1,\beta_1} &A_{m-1}\end{bmatrix}-
  \sum\limits_{\tiny\begin{matrix}Q_{1j}, Q_{2j}\in \mathcal{P}_M\\j\in[m]\end{matrix}}
  \perm\begin{bmatrix}  
    Q_{11,\beta_1}& A_1\\ ~\\
\left.\begin{matrix}\begin{matrix}0 &1\\1&0\\\vdots &\vdots \\0&0\end{matrix}&\cdots&\end{matrix}\right|&A_2\\
\vdots& \vdots\\
Q_{m-1,1,\beta_1} &A_{m-1}\end{bmatrix}\label{formula1}\\\nonumber ~\\=
 &\sum\limits_{\tiny\begin{matrix}Q_{1j}, Q_{2j}\in \mathcal{P}_M\\j\in[m]\end{matrix}}
 \perm\begin{bmatrix}  
   Q_{11,\beta_1\setminus \{1\}}& A_1\\ ~\\
\left. \begin{matrix} \begin{matrix}0\\0\\\vdots \\0\end{matrix}&\cdots& \end{matrix}\right| &A_2{\setminus {\rm row} 1}\\
 \vdots & \vdots \\
Q_{m-1,1,\beta_1\setminus \{1\}} &A_{m-1}\end{bmatrix}+
  \sum\limits_{\tiny\begin{matrix}Q_{1j}, Q_{2j}\in \mathcal{P}_M\\j\in[m]\end{matrix}}
  \perm\begin{bmatrix}  
  Q_{11,\beta_1\setminus \{2\}}& A_1\\ ~\\
 \left.\begin{matrix} \begin{matrix}0 \\0\\\vdots \\0\end{matrix} &\cdots&\end{matrix}\right| &A_2{\setminus {\rm row} 2}\\
 \vdots&\vdots\\
  Q_{m-1,1,\beta_1\setminus \{2\}} &A_{m-1}\end{bmatrix} \nonumber\\-&\sum\limits_{\tiny\begin{matrix}Q_{1j}, Q_{2j}\in \mathcal{P}_M\\j\in[m]\end{matrix}}
 \perm\begin{bmatrix}  
  Q_{11,\beta_1\setminus \{1\}}& A_1\\ ~\\
 \left.\begin{matrix} \begin{matrix}0\\0\\\vdots \\0\end{matrix} &\cdots &\end{matrix}\right| &A_2{\setminus {\rm row} 2}\\
 \vdots & \vdots\\
 Q_{m-1,1,\beta_1\setminus \{1\}} &A_{m-1}\end{bmatrix}-
  \sum\limits_{\tiny\begin{matrix}Q_{1j}, Q_{2j}\in \mathcal{P}_M\\j\in[m]\end{matrix}}
  \perm\begin{bmatrix}  
Q_{11,\beta_1\setminus \{2\}}& A_1\\ ~\\
\left.\begin{matrix}  \begin{matrix}0 \\0\\\vdots \\0\end{matrix} &\cdots &\end{matrix}\right|&{A}_2{\setminus {\rm  row} 1}\\
\vdots& \vdots\\
 Q_{m-1,1,\beta_1\setminus \{2\}} &A_{m-1}\end{bmatrix}\label{formula2}=0
  \end{align}}
\addtocounter{equation}{2} \\each $P_{ij}$ is a permutation  matrix of size $M\times M$, and   $T_\beta$ has the same permanent as the latter matrix. 
 \begin{example} \label{example2}
 Let  $T_\beta$ with constant row weight $m-1=2$  as in \eqref{matrixexample1} and let $\setS_{12},  \setS_{13}, \setS_{22},  \setS_{23}$, as in \eqref{matrixexample2}. (Note that $T_\beta$  is the matrix  $T''_\beta$ in Example \ref{example1}.)
Let $$P_{11}\defeq \begin{bmatrix}Q_{12,\beta_2}&\setS_{12} \end{bmatrix},\quad  P_{12}\defeq \begin{bmatrix}Q_{13,\beta_3}&\setS_{13} \end{bmatrix},$$ $$P_{21}\defeq \begin{bmatrix}Q_{22,\beta_2}&\setS_{22} \end{bmatrix},\quad
P_{22}\defeq \begin{bmatrix}Q_{23,\beta_3}&\setS_{23} \end{bmatrix}, $$ as described in \eqref{permutationsexample2}. 
 
Let $\sigma$ denote the permutation operator  that permutes the columns of $T_\beta$ such that the first column gets moved  to the 6th position and the second column gets moved to the 3rd position, thus completing the matrices $ Q_{12,\beta_2}$ and $Q_{13,\beta_3}$ to the permutation matrices   $P_{11}$ and $P_{12} $,  respectively. 
We compute $\sigma(T_\beta)$ and obtain $$\sigma(T_\beta)=\begin{bmatrix} \begin{matrix} 0&1&0\\1&0&0\\0&0&1\end{matrix}&~&\begin{matrix} 0&0&1
 \\0&1&0\\1&0&0\end{matrix}\\ ~ \\ \begin{matrix} 1&0&0\\0&0&1\\0&1&0\end{matrix}&~&\begin{matrix} 1&0&0\\0&1&0\\0&0&1\end{matrix}\end{bmatrix}\defeq \begin{bmatrix}P_{11}&P_{12}\\P_{21}&P_{22}\end{bmatrix},$$ since the $\sigma$ column permutation completes also the matrices  $ Q_{22,\beta_2}$ and $Q_{23,\beta_3}$ to the permutation matrices   $P_{21}$ and $P_{22} $,  respectively. This achieves the desired form. 
 \end{example}
 
  We will now address the case where the order does not get preserved in the next $M$ rows. For simplicity, we suppose that $\setS_{2j}\neq\setS_{1j}$. Therefore, we need an extra permutation of columns within the block matrix $Q_{21,\beta_1}$ so that we obtain the equality $\setS_{2j}=\setS_{1j}$. In addition, we need to show that this change does not alter the sum of the permanents of all  $\sigma(T_\beta)$ 
 over all possible permutation matrices $Q_{ij}$.
 In fact, it is sufficient to show that the operation of interchanging two columns within the matrix $Q_{21,\beta_1}$ does not change the sum of $\perm(T_\beta)$ when computed over all possible permutation matrices $Q_{ij}$. This fact, applied sequentially to each matrix  $Q_{i1,\beta_1}$, $i\geq 2$, until the desired form is obtained,  will conclude our proof.  
 
 Without loss of generality, for simplicity, we can assume that the two columns of $Q_{21,\beta_1}$ that get interchanged  are the vectors $\begin{bmatrix} 1&0&0&\cdots&0\end{bmatrix}^\tr$ and $\begin{bmatrix} 0&1&0&\cdots&0\end{bmatrix}^\tr$. Let  us denote by $Q'_{21,\beta_1}$  the matrix obtained from $Q_{21,\beta_1}$, and $T'_\beta$ be the matrix obtained from $T_\beta$, after the two columns get interchanged.  In equation  \eqref{formula1} we compute the difference of the permanents of the two matrices and sum this difference over all permutation matrices $Q_{1j}$ and $Q_{2j}$. We denote $A_i \defeq \begin{bmatrix}Q_{i2,\beta_2}&\cdots &Q_{i, m-1, \beta_m}\end{bmatrix}$, for all $i\in [m-1]$. After performing consecutive cofactor expansions on the first and second row of the second block of $M$ rows of $T_\beta$, we obtain that this difference is given by \eqref{formula2}, where $A_2\setminus{\rm row1}$ denotes the submatrix of $A_2$ obtained by erasing its first row and $A_2\setminus{\rm row2}$ denotes the  submatrix of $A_2$ obtained by erasing its second row.    We note that the first and the last sums of the four sums  in \eqref{formula2} are equal, and so are the second and the third sums, and so the four sums cancel each other and give $$\sum\limits_{Q_{1j}, Q_{2j}\in \mathcal{P}_M}\perm(T_\beta)-\perm(T'_\beta)=0.$$

Summing over all matrices $Q_{ij}\in \mathcal{P}_M$, we obtain 
\begin{align*}&\sum\limits_{\tiny\begin{matrix}Q_{ij}\in \mathcal{P}_M\\ i,j\in [M]\end{matrix}}\perm(T_\beta)-\perm(T'_\beta)=\end{align*}
\begin{align*}
=& \sum\limits_{\tiny\begin{matrix}Q_{ij}\in \mathcal{P}_M\\ i\in [M]\setminus \{1,2\}\end{matrix}}\sum\limits_{\tiny\begin{matrix}Q_{1j}, Q_{2j}\in \mathcal{P}_M\\ j\in [M]\end{matrix}}\perm(T_\beta)-\perm(T'_\beta)=0.\end{align*}
 Therefore we can change the position of the columns within each matrix $Q_{i1,\beta_1}$, for all $i\geq 2$, until we obtain a matrix  of the form given in \eqref{smaller}, without changing the sum of the permanents of $T_\beta$ computed over all permutation matrices $Q_{ij}$. This, together with the previous case, concludes the proof. 
\end{IEEEproof}\begin{example}\label{example3}
Let  
 \begin{align*} T_\beta&\defeq \begin{bmatrix} Q_{11,\beta_1}& Q_{12,\beta_2}&Q_{13,\beta_3}\\Q_{21,\beta_1}& Q_{22,\beta_2}&Q_{23,\beta_3}\end{bmatrix}\\&\defeq \begin{bmatrix} \begin{matrix} 1&0\\0&0\\0&1\end{matrix}&~&\begin{matrix} 0&1\\1&0\\0&0\end{matrix}&~&\begin{matrix} 0&0 \\0&1\\1&0\end{matrix}\\ ~ \\ \begin{matrix} 0&0\\1&0\\0&1\end{matrix}&~&\begin{matrix} 1&0\\0&0\\0&1\end{matrix}&~&\begin{matrix} 1&0\\0&1\\0&0\end{matrix}\end{bmatrix} \end{align*} with constant row weight $m-1=2$.   Let $$\setS_{12}=\begin{bmatrix} 0\\0\\1\end{bmatrix}, \quad \setS_{13}=\begin{bmatrix} 1\\0\\0\end{bmatrix}, \setS_{22}=\begin{bmatrix} 0\\1\\0\end{bmatrix}, \quad \setS_{23}=\begin{bmatrix} 0\\0\\1\end{bmatrix}$$
and $$P_{11}\defeq \begin{bmatrix}Q_{12,\beta_2}&\setS_{12} \end{bmatrix}=\begin{bmatrix} 0&1&0\\1&0&0\\0&0&1\end{bmatrix},$$
 $$P_{12}\defeq \begin{bmatrix}Q_{13,\beta_3}&\setS_{13} \end{bmatrix}=\begin{bmatrix}0&0&1\\0&1&0\\1&0&0\end{bmatrix},$$
  $$P_{21}\defeq \begin{bmatrix}Q_{22,\beta_2}&\setS_{22} \end{bmatrix}=\begin{bmatrix} 1&0&0\\0&0&1\\0&1&0\end{bmatrix},$$
$$P_{22}\defeq \begin{bmatrix}Q_{23,\beta_3}&\setS_{23} \end{bmatrix}=\begin{bmatrix}1&0&0\\0&1&0\\0&0&1\end{bmatrix}.$$

Let $\sigma$ denote the permutation operator  that permutes the columns of $T_\beta$ such that the first column gets moved  to the 6th position and the second column gets moved to the 3rd position, thus completing the matrices $ Q_{12,\beta_2}$ and $Q_{13,\beta_3}$ to the permutation matrices   $P_{11}$ and $P_{12} $,  respectively. 
We compute $\sigma(T_\beta)$ and obtain $$\sigma(T_\beta)=\begin{bmatrix}P_{11}&P_{12}\\R_{21}&R_{22}\end{bmatrix}=\begin{bmatrix} \begin{matrix} 0&1&0\\1&0&0\\0&0&1\end{matrix}&~&\begin{matrix} 0&0&1
 \\0&1&0\\1&0&0\end{matrix}\\ ~ \\ \begin{matrix} 1&0&0\\0&0&0\\0&1&1\end{matrix}&~&\begin{matrix} 1&0&0\\0&1&1\\0&0&0\end{matrix}\end{bmatrix}. $$ The  $\sigma$ column permutation does not complete  the matrices  $ Q_{22,\beta_2}$ and $Q_{23,\beta_3}$ to the permutation matrices   $P_{21}$ and $P_{22} $,  respectively. 

Let $$T'_\beta\defeq \begin{bmatrix} \begin{matrix} 0&1&0\\1&0&0\\0&0&1\end{matrix}&~&\begin{matrix} 0&0&1
 \\0&1&0\\1&0&0\end{matrix}\\ ~ \\ \begin{matrix} 1&0&0\\0&0&1\\0&1&0\end{matrix}&~&\begin{matrix} 1&0&0\\0&1&0\\0&0&1\end{matrix}\end{bmatrix}= \begin{bmatrix}P_{11}&P_{12}\\P_{21}&P_{22}\end{bmatrix}$$
obtained by interchanging the 3rd and 6th columns of $\sigma(T_\beta)$. The above computations show that  this change does not alter the average permutation over all permutation matrices $\matr{Q}\in \Psi_{2,3,3}$.  $T'_\beta$ has now the desired form.
\end{example}

\subsection{Exact Formula for $q_{2,M}$}\label{exact:q2} 
\begin{theorem}\label{2-3case} $q_{2,M}=M+1. $ 
\end{theorem} 
\begin{IEEEproof} 
 Using the technique from Lemma \ref{recursive}, we compute $q_{2,M} $ as follows: 
\begin{align*}q_{2,M}=&{{\sum\limits_{P, Q, R, S\in \mathcal{P}_M}\perm\begin{bmatrix} P &Q\\ R& S
\end{bmatrix}}\over{(M!)^4}}={{\sum\limits_{P\in \mathcal{P}_M}\perm\begin{bmatrix} I &I\\ I& P
\end{bmatrix}}\over{M!}}\end{align*}

Therefore, we need to compute $\perm\begin{bmatrix} I &I\\ I& P\end{bmatrix}$, for some  $P\in \mathcal{P}_M$ and then to average over all possible  $P\in \mathcal{P}_M$. 
Since the permanent is computed by summing all products of entries such that each row and each column contributes to the product exactly once, the first $M$ rows of $ \begin{bmatrix} I &I\\ I& P\end{bmatrix}$ can contribute with 1s from the two adjacent identity matrices (on top)  by choosing  the 1 entries from the set of  columns  
$[M]\setminus \beta_1$ with $r\defeq \left|\beta_1\right|$,  $0\leq r\leq M$,
  from the first block, and $\beta_1$  from the second block.  This implies  that  the columns indexed by $[M]\setminus \beta_1$ 
  from the first block and $\beta_1$ from the second block cannot contribute anymore to the rest of the product entries. Given $r$, there are  $ {M \choose M-r}={M \choose r}$ ways of choosing the $M-r$ columns in $[M]\setminus \beta_1$.  Therefore, we obtain    
  \begin{align*}&\!\!\!\!\!\perm\begin{bmatrix} I &I\\ I& P
\end{bmatrix}=\sum\limits_{r=0}^{M} {M \choose r}\!\!\!\!\sum\limits_{\tiny\begin{matrix}\beta_1\subseteq [M]\\|\supp(\beta_1)|=r\end{matrix}}\perm\begin{bmatrix}  I_{\beta_1}&P_{[M]\setminus \beta_1}\end{bmatrix}. 
\end{align*} 
  Then
 \begin{align*}q_{2,M}=&\sum\limits_{r=0}^{M} {M \choose r}\sum\limits_{\tiny\begin{matrix}\beta_1\subseteq [M]\\|\supp(\beta_1)|=r\end{matrix}}\sum\limits_{{P}\in \mathcal{P}_M}\perm\begin{bmatrix}  I_{\beta_1}& P_{[M]\setminus \beta_1}\end{bmatrix}.
\end{align*} 
The matrix $\begin{bmatrix}  I_{\beta_1}& P_{[M]\setminus \beta_1}\end{bmatrix}$ is an $M\times M$ matrix with $r$  of its columns indexed by $\beta_1$  coming from the identity matrix, and the rest from the permutation matrix $P$. If two columns in $I_{\beta_1}$ and $ P_{[M]\setminus \beta_1}$ are equal, then the matrix must have a zero row, which in turn, gives a zero permanent. Zero permanents do not count in the sum, so we need to count how many permutation matrices $P$ give a non-zero permanent  $\perm\begin{bmatrix}  I_{\beta_1}& P_{[M]\setminus \beta_1}\end{bmatrix}$. For a fixed $\beta_1$,  this permanent is non-zero if the columns of $I$ indexed by $\beta_1$  are a permutation of the columns of $P$ indexed by $\beta_1$, (or, equivalently, $\begin{bmatrix}  I_{\beta_1}& P_{[M]\setminus \beta_1}\end{bmatrix}$ is a permutation matrix) in which case the permanent is 1. 
For a fixed $\beta_1$, there are $(M-r)!$ ways  of choosing the $M-r$ columns of $P$  and $r!$ ways  of choosing the remaining columns of $P$ to obtain all possible matrices $P\in \mathcal{P}_M $ that give a permanent 1 for $\perm\begin{bmatrix}  I_{\beta_1}& P_{[M]\setminus \beta_1}\end{bmatrix}$.    We obtain:
\begin{align*}q_{2,M}=&{{ \sum\limits_{r=0}^{M} {M \choose r} r! (M-r)!\cdot 1}\over{M!}}=M+1.
\end{align*} 
\end{IEEEproof} 

\subsection{Exact Formulas for ${t}_{3,M}$ and  ${\hat t}_{3,M}$}\label{exact:t3}
Let \begin{align} \label{formula:t} t_{3,M} &\defeq {\left<\perm\begin{bmatrix}1&1&0\\1&1&1\\1&1&1\end{bmatrix}^{\uparrow\matr{P}}\right>_{\matr{P}\in \Psi_{3,M}}}\end{align}
and \begin{align} {\hat t}_{3,M}&\defeq {\left<\perm\begin{bmatrix}1&1&0\\1&0&1\\1&1&1\end{bmatrix}^{\uparrow\matr{P}}\right>_{\matr{P}\in \Psi_{3,M}}}.\label{formula:that}\end{align}
\begin{theorem}\label{3-3-0case} The following are exact formulas for $t_{3,M}$ and ${\hat t}_{3,M}$.\\
$t_{3,M}={{ \sum\limits_{r=0}^{M} {M \choose r} \sum\limits_{s=0}^{r} {r \choose s}(M-r+s)!(M-s)! \sum\limits_{t=0}^{M-r} {M-r \choose t}(M-t)! (r+t)!}\over{M!^3}}.$
${\hat t}_{3,M}={{ \sum\limits_{r=0}^{M} {M \choose r} \sum\limits_{s=0}^{r} {r \choose s}(M-r+s)!(M-s)!  r!}\over{M!^2}}.$
\end{theorem} 
\begin{IEEEproof}
  Applying Lemmas \ref{lemma11} and \ref{theorem3} in Appendix  \ref{lemmas}, we can compute the permanent $\perm\left(\begin{matrix} I &I&0\\ I& P&I\\Q&R&I
 \end{matrix}\right)$ as in \eqref{formulatopsum} on top of the next page. 
 The permanents $$\perm \begin{bmatrix} I_{\gamma_1}&P_{\alpha_1\setminus \gamma_1}&Q_{\gamma_2}&R_{[M]\setminus \alpha_1\setminus\gamma_2} \end{bmatrix}$$ in \eqref{formulatopsum} are 0  unless  $ \begin{bmatrix}I_{\gamma_1}&P_{\alpha_1\setminus \gamma_1}&Q_{\gamma_2}&R_{[M]\setminus \alpha_1\setminus\gamma_2} \end{bmatrix}$ is an  $M\times M$  permutation matrix.  
Then, taking the  average permanent over all permutation matrices $P,Q,R$, yields \\$ t_{3,M}=
 {{ \sum\limits_{r=0}^{M}\!\! {M \choose r} \!\!\sum\limits_{s=0}^{r} \!\!{r \choose s}\!\!\sum\limits_{t=0}^{M-r} \!\!{M-r \choose t} (M-r+s)!(M-s)! (M-t)! (r+t)!}\over{M!^3}}, $  where 
${0\choose 0}=1$ by definition. 
  
  Then, ${\hat t}_{3,M}$ can be computed by taking $t=0$ in the ${t}_{3,M}$ formula.
 \end{IEEEproof}  

\begin{remark}  
The above formulas for the average can  also be rewritten as
\begin{align*}
t_{3,M} 
 &= \sum\limits_{r=0}^{M} {{\sum\limits_{s=0}^{r} {M-r+s \choose M-r}{M-s \choose M-r}
 \sum\limits_{t=0}^{M-r} {M-t \choose r}{r+t\choose r}} \over { {M\choose r}^2}}=\\
 &=\sum\limits_{r=0}^{M}{{{2M-r+1 \choose 2M-2r+1} {M+r+1 \choose 2r+1}}\over {{M\choose r}^2}} \; .
  \end{align*}
  Taking $t=0$, we obtain
\begin{align*}  {\hat t}_{3,M} 
 &= \sum\limits_{r=0}^{M} {{\sum\limits_{s=0}^{r} {M-r+s \choose M-r}{M-s \choose M-r}} \over { {M\choose r}}}=\sum\limits_{r=0}^{M}{{{2M-r+1 \choose 2M-2r+1} }\over {{M\choose r}}} \; .
 \end{align*}
 \end{remark}
\subsection{Necessary Lemmas}\label{lemmas}
 This Appendix contains a few results that we need in our computations. Versions of the first two lemmas can also be found in \cite{Minc:78}.
\begin{lemma}\label{split} Let $\vect{c}_i\in \Ftwo^{(M-1)\times 1}$, $\matr{A}\in \Ftwo^{(M-1)\times (M-2)}$ and the all-zero matrix $\zero $ of size $1 \times (M-2)$. Then, the following ``expansion" of the permanent holds:

\begin{align}\nonumber
\perm \begin{bmatrix} 1&1 &{\zero}\\ \matr{c}_1&\matr{c}_2 &\matr{A}
\end{bmatrix}=  \perm \begin{bmatrix} \matr{c}_1+\matr{c}_2 &\matr{A}
\end{bmatrix}.\end{align} \end{lemma}
 \begin{IEEEproof} Applying the cofactor expansion on the first row, the above permanent can be rewritten, based on the row-linear property of both determinant and permanent, as 

 $\perm \begin{bmatrix} \matr{c}_2 &\matr{A}
\end{bmatrix}+\perm \begin{bmatrix}\matr{c}_1 &\matr{A}\end{bmatrix}=  \perm \begin{bmatrix} \vect{c}_1+\vect{c}_2 &\matr{A} \end{bmatrix}.$
\end{IEEEproof}

 \begin{lemma}\label{lemma11}
 Let $\matr{A}, \matr{B}$ be two square matrices of the same size $M$. Then
\begin{align}\label{matrixsplit2}\perm (\matr{A}+\matr{B})= \sum _{\alpha\subseteq[M]}\perm\begin{bmatrix} 
\matr{A}_\alpha& \matr{B}_{[M]\setminus \alpha}
\end{bmatrix}.\end{align} 
\end{lemma}
\begin{IEEEproof} The permanent of the matrix is a sum of all possible non-zero elementary products. Each elementary product contains a 1 entry from each row and each column and this entry can be taken either from the matrix $\matr{A}$ or $\matr{B}$, giving the above sum.  \end{IEEEproof}
Using these two lemmas the following result can be derived.
\begin{lemma}\label{theorem3}
 Let $P,Q,R,S$ be permutation matrices, and $I$ the identity matrix, all of size $M$.
 Then
 \begin{align}
 \perm\begin{bmatrix} I &I&0\\ I& P&I\\Q&R&I
\end{bmatrix}=\perm(I+P+Q+R).
\end{align}
\end{lemma}
\begin{IEEEproof}We apply Lemma \ref{split} sequentially for the first $M$ rows and then the last $M$ columns, and we get 
\begin{align*}\perm\begin{bmatrix} I &I&0\\ I& P&I\\Q&R&I
\end{bmatrix}=&\perm\begin{bmatrix} I+P&I \\Q+R& I\end{bmatrix}\\=& \perm(I+P+Q+R). \\[-1.2cm]\end{align*}
\end{IEEEproof}

 \bigformulatop{\value{equation}}
{\begin{align}&\perm\left(\begin{matrix} I &I&0\\ I& P&I\\Q&R&I
 \end{matrix}\right)=\perm(I+P+Q+R)\nonumber\\=&\sum\limits_{r=0}^{M} \sum\limits_{\tiny\begin{matrix}\alpha_1\subseteq [M]\\|\supp(\alpha_1)|=r\end{matrix}}\perm\begin{bmatrix} (I+P)_{\alpha_1} &(Q+R)_{[M]\setminus \alpha_1}\end{bmatrix}\nonumber\\ =&
 \sum\limits_{r=0}^{M}\!\!\!  \sum\limits_{\tiny\begin{matrix}\alpha_1\subseteq [M]\\|\supp(\alpha_1)|=r\end{matrix}}  \sum\limits_{s=0}^{r} \!\!\!  \sum\limits_{\tiny\begin{matrix}\gamma_1\subseteq \alpha_1\\|\supp(\gamma_1)|=s\end{matrix}}\!\!\! \sum\limits_{t=0}^{M-r}  \sum\limits_{\tiny\begin{matrix}\gamma_2\subseteq [M]\setminus \alpha_1\\|\supp(\gamma_2)|=t\end{matrix}}\!\!\! \!\!\!\!\!\perm
\begin{bmatrix} I_{\gamma_1}&P_{\alpha_1\setminus \gamma_1}&Q_{\gamma_2}&R_{[M]\setminus \alpha_1\setminus\gamma_2} \end{bmatrix}\label{formulatopsum}
\end{align}}
\addtocounter{equation}{0}

\bibliographystyle{ieeetr}
 \bibliography{../Bib-Files/huge.bib} 
\end{document}